\documentclass[epj]{svjour}

\usepackage[intlimits]{amsmath}
\usepackage{amsfonts}
\usepackage{amssymb,amscd}
\usepackage{color}
\usepackage[dvips]{epsfig}

\newcommand*{\no}{\noindent}
\newcommand*{\bea}{\begin{eqnarray}}
\newcommand*{\eea}{\end{eqnarray}}
\newcommand*{\be}{\begin{equation}}
\newcommand*{\ee}{\end{equation}}
\newcommand*{\pd}{\partial}

\newcommand*{\mn}{{\mu\nu}}

\newcommand*{\nn}{\nonumber}

\newcommand{\beq}{\begin{equation}}
\newcommand{\eeq}{\end{equation}}
\newcommand{\beqa}{\begin{eqnarray}}
\newcommand{\eeqa}{\end{eqnarray}}

\newcommand{\vp}{\vec{p}}
\newcommand{\vq}{\vec{q}}
\newcommand{\vk}{\vec{k}}
\newcommand{\op}{\omega_{p}}
\newcommand{\oq}{\omega_{q}}
\newcommand{\ok}{\omega_{k}}

\newcommand{\intk}{\sum_{n_k} \int \frac{d^3k}{(2\pi)^3}}

\def\Eq#1{Eq.~(\ref{#1})}

\begin{document}

\title{Chiral and deconfinement transition from correlation functions:\\ SU(2) vs.\ SU(3)}

\author{Christian~S.~Fischer\inst{1,2}
\and Axel Maas
\inst{3}
\thanks{\emph{e-mail:} axelmaas@web.de}
\and Jens A.\ M\"uller
\inst{1}
\thanks{\emph{e-mail:} jens.mueller@physik.tu-darmstadt.de
}}

\institute{ Institut f\"ur Kernphysik, Technische Universit\"at Darmstadt,
            Schlossgartenstra{\ss}e 9, D-64289 Darmstadt, Germany \and
	    GSI Helmholtzzentrum f\"ur Schwerionenforschung GmbH, 
            Planckstr. 1  D-64291 Darmstadt, Germany \and
	    Institut f\"ur Physik, Karl-Franzens Universit\"at Graz, 
	    Universit\"atsplatz 5, A-8010 Graz, Austria}

\date{Received: date / Revised version: date}
\abstract{
We study a gauge invariant order parameter for deconfinement and the chiral 
condensate in SU(2) and SU(3) Yang-Mills theory in the vicinity of the 
deconfinement phase transition using the Landau gauge quark and gluon 
propagators. We determine the gluon propagator from lattice calculations 
and the quark propagator from its Dyson-Schwinger equation, using the gluon 
propagator as input. The critical temperature and a deconfinement order 
parameter are extracted from the gluon propagator and from the dependency 
of the quark propagator on the temporal boundary conditions. The chiral
transition is determined using the quark condensate as order parameter.
We investigate whether and how a difference in the chiral and deconfinement
transition between SU(2) and SU(3) is manifest.
}
\maketitle
\newpage

\section{Introduction}

Two of the most characteristic features of QCD are at the same time two of
the most elusive ones: Confinement and chiral symmetry breaking. Of particular 
interest is the dependency of these phenomena on temperature and density. 
For vanishing current quark masses chiral symmetry gets restored by a phase transition
above a critical temperature. On the other hand in the limit of infinitely heavy quarks 
a phase transition from a confining phase to a deconfined phase takes place. Of course, 
in full QCD with physical quarks none of the aforementioned limits is appropriate. 
Indeed, at zero chemical potential the would-be order parameters for both transitions 
show a rapid change signaling a well-marked cross-over region \cite{bazavovaoki}.
Furthermore it seems remarkable
that the temperature ranges for both cross-overs
are notable close to each other \cite{Cheng:2009zi,Cheng:2009be,Aoki:2009sc}.
This is in contrast to {\it e.g.} the case of adjoint quarks 
where both temperatures differ by a factor of almost eight \cite{Luetgemeier98}.

Thus, it warrants to investigate the chiral and deconfinement transition for 
fundamental quarks closely, to eventually understand both properties. To avoid 
the problems involved with dynamical quarks, a useful laboratory is the 
quenched approximation where only valence quarks are present. In this case the 
deconfinement transition can be cleanly defined by usage of the Polyakov-loop 
as order parameter, and both the chiral restoration and deconfinement temperature 
coincide within available accuracy \cite{Karsch:2003jg}.

Useful tools to address these questions are the so-called dual observables
\cite{dualcond1,Ga1,Jena1,BiGa,G2,Su2adj,Fischer:2009wc,Fischer:2009gk,Bilgici:2009tx,soeldner,Braun:2009gm}.
These are obtained by modifying the temporal boundary conditions 
of the valence (test) 
quarks, without altering the dynamics of the theory\footnote{Such an alteration 
would correspond to the introduction of an imaginary chemical potential 
\cite{Roberge:1986mm} as discussed in Ref.~\cite{Braun:2009gm}.}. 
Dual observables are sensitive to the spectral 
properties of the Dirac operator, and thus do encode both the confinement 
and the chiral properties of quarks. In particular, they are not only able to 
distinguish the low-temperature from the high-temperature phase, but 
also phases where chiral symmetry is broken but confinement is no 
longer present \cite{Su2adj}.

Dual observables have been first defined using lattice gauge theory \cite{Ga1}, 
but have turned out to be quite expensive to determine. In particular, the large
distances characteristic of confining physics and the small masses relevant to 
chiral symmetry are hard to reach. On the other hand, it is also possible to 
determine dual quantities using functional methods \cite{Fischer:2009wc,Braun:2009gm}. 
In this case, there is no limit neither on distance nor on the quark masses. 
The challenge is, however, that truncations are necessary. In this work we combine the 
best of both worlds by a combination of lattice and continuum methods 
\cite{Fischer:2009wc,Fischer:2009gk,Cucchieri:2007ta}.

The goal of our study is twofold. On the one hand we investigate (potentially gauge-dependent) 
mechanisms linking deconfinement and chiral symmetry restoration. The study  
presented in this work in particular highlights the role of the longitudinal, electric
part of the Landau gauge gluon propagator in this respect. On the other hand our study serves as an
important intermediate step towards an analysis of the QCD phase diagram at non-vanishing
chemical potential. Due to the notorious sign problem of lattice QCD in that realm there
is great demand for other methods beyond model calculations. Functional methods like
Dyson-Schwinger equations \cite{Alkofer:2000wg,Roberts:2000aa} and the functional 
renormalization group \cite{Schaefer:2006sr} are ready to fill this gap. To provide for
meaningful results at non-zero chemical potential it is, however, first necessary to gain
as much insight as possible into the fidelity of truncations at zero chemical potential,
where one can compare with the results from lattice QCD. As a result of this work we
find a clear (gauge-invariant) signal for the deconfinement transition from the dressed 
Polyakov-loop determined from the Dyson-Schwinger equations in agreement with the information
we extract from the lattice gluon propagator.

This paper is organised as follows: Our method to determine the magnetic and electric 
part of the Landau gauge gluon propagator on the lattice are discussed in section 
\ref{sec:lattice}. Though the gluon propagator is gauge-dependent, it is found that 
it can be used to obtain gauge-invariant properties, like the critical temperature.
The gluon propagator is also one of the key ingredients into the Dyson-Schwinger equation
for the quark propagator. In principle the gluon propagator could be obtained using 
Dyson-Schwinger equations \cite{Cucchieri:2007ta,Zahed:1999tg,Maas:2005hs,Maas:2004se}, 
but this turns out to be a formidable task \cite{Cucchieri:2007ta}. Instead we use
our result from lattice gauge theory as input, as described in section \ref{sec:dse}. 
From the Landau gauge quark propagator we then extract the gauge-invariant order
parameters for the chiral and deconfinement transition. 
This procedure has been used previously in \cite{Fischer:2009wc,Fischer:2009gk}. 
However, the then available lattice data \cite{Cucchieri:2007ta} for the gluon 
propagator have been very coarse on the temperature axis, and only available for 
SU(2). We expand here to a much finer grid in the temperature domain, while at 
the same time employing much larger physical volumes on the lattice. Preliminary 
results of these calculations can be found in \cite{Maas:2009fg}. In addition, we 
study the physical case of SU(3)\footnote{First explorative studies of the gluon 
propagator in the unquenched case can be found in \cite{Furui:2006py}.}. The results
are summarized in section \ref{sec:sum}.

\begin{figure*}[Htb]
\includegraphics[width=0.4\linewidth]{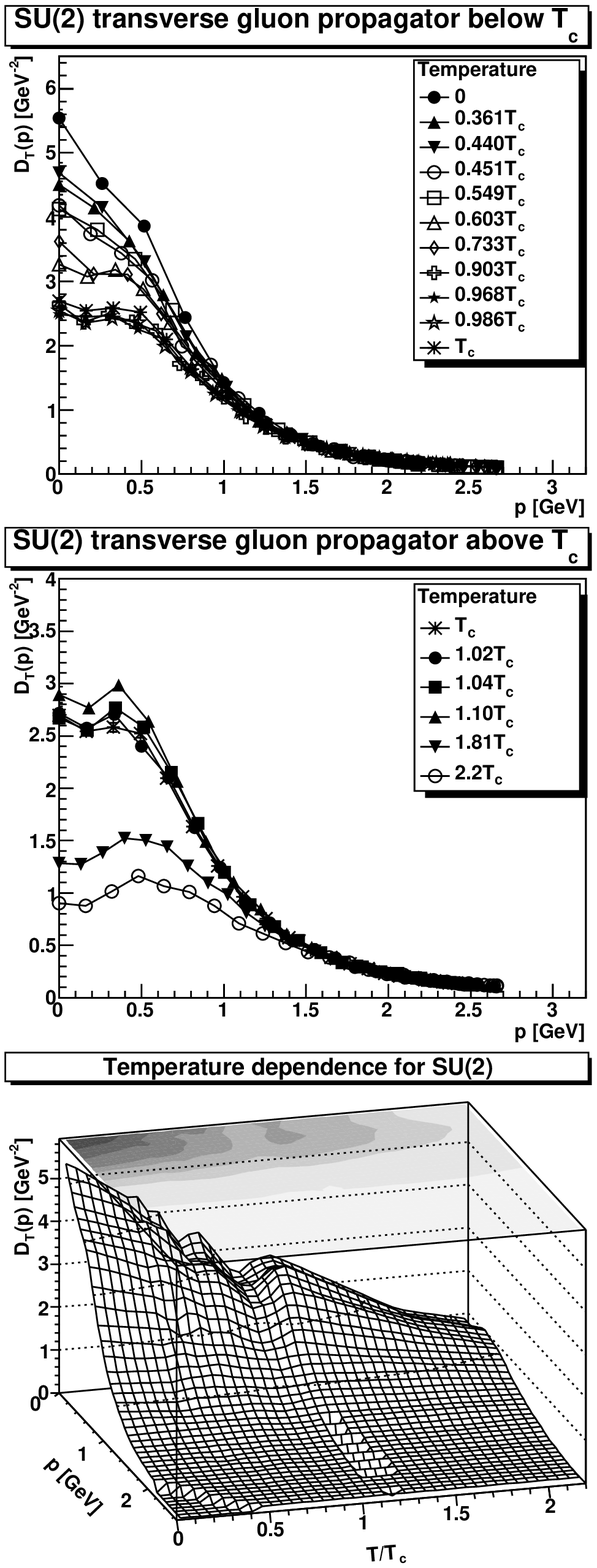}\hfill
\includegraphics[width=0.4\linewidth]{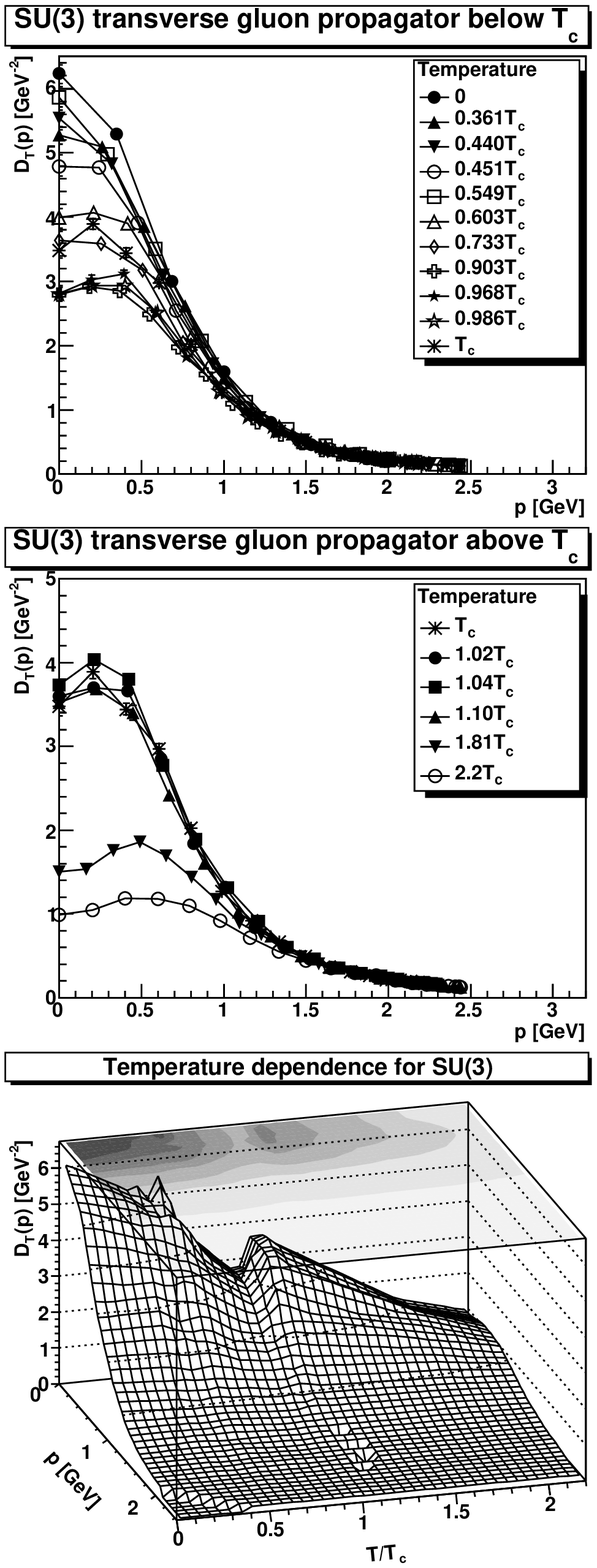}
\caption{\label{lat-gpt}The (soft mode of the) transverse part $D_T$ of the gluon propagator at 
temperatures below (top panel) and above (middle panel) the phase transition. 
The bottom panel shows both results together. On the left results for SU(2) and 
on the right results for SU(3) are shown. Lines are drawn to guide the eye. 
Momenta here and hereafter are aligned along the $x$-axis.}
\end{figure*}

\begin{figure*}[htb]
\includegraphics[width=0.4\linewidth]{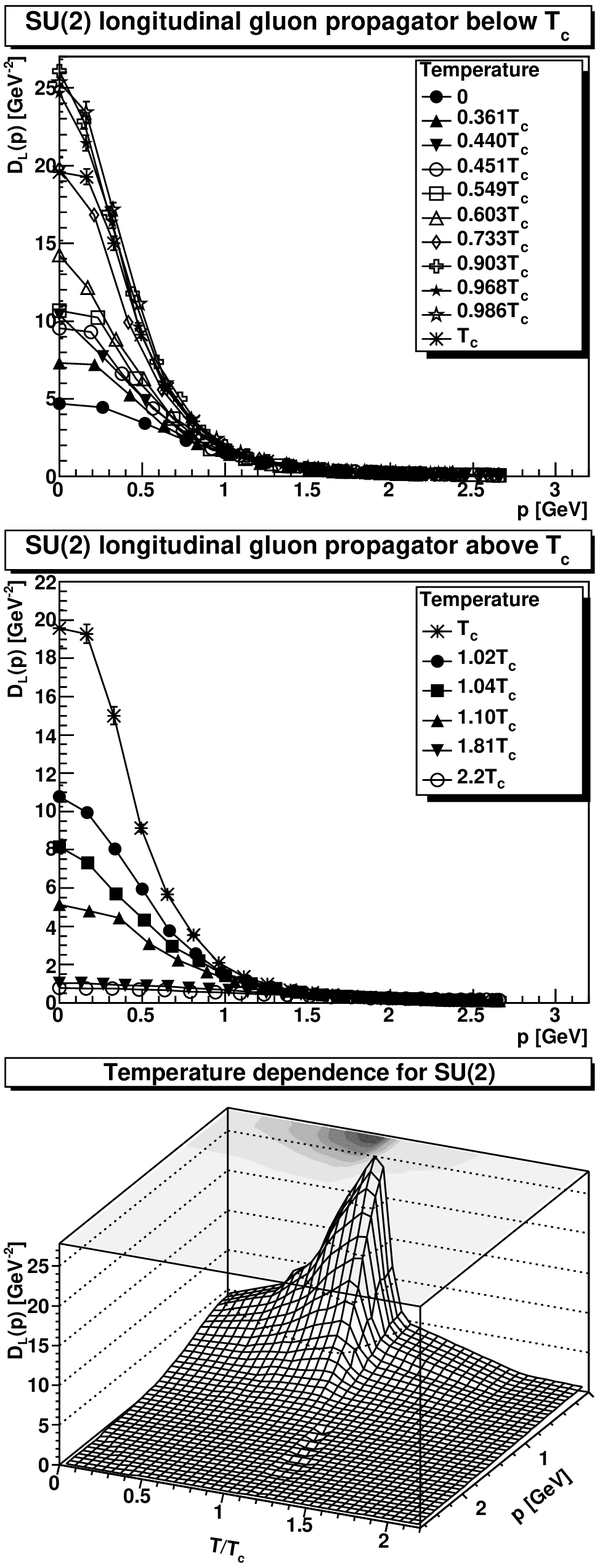}\hfill
\includegraphics[width=0.4\linewidth]{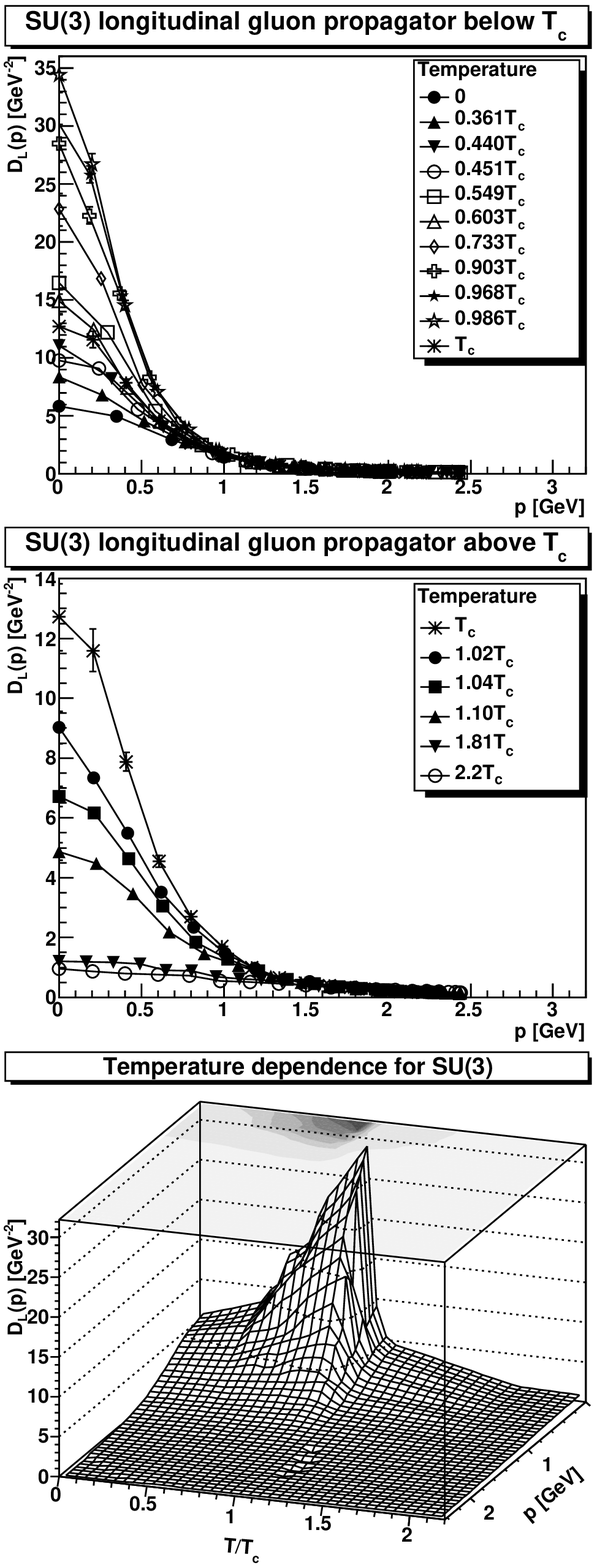}
\caption{\label{lat-gpl}The (soft mode of the) longitudinal part $D_L$ of the gluon propagator at 
temperatures below (top panel) and above (middle panel) the phase transition. 
The bottom panel shows both results together. On the left results for SU(2) 
and on the right results for SU(3) are shown. Lines are drawn to guide the eye.}
\end{figure*}

\begin{figure*}[htb]
\includegraphics[width=0.4\linewidth]{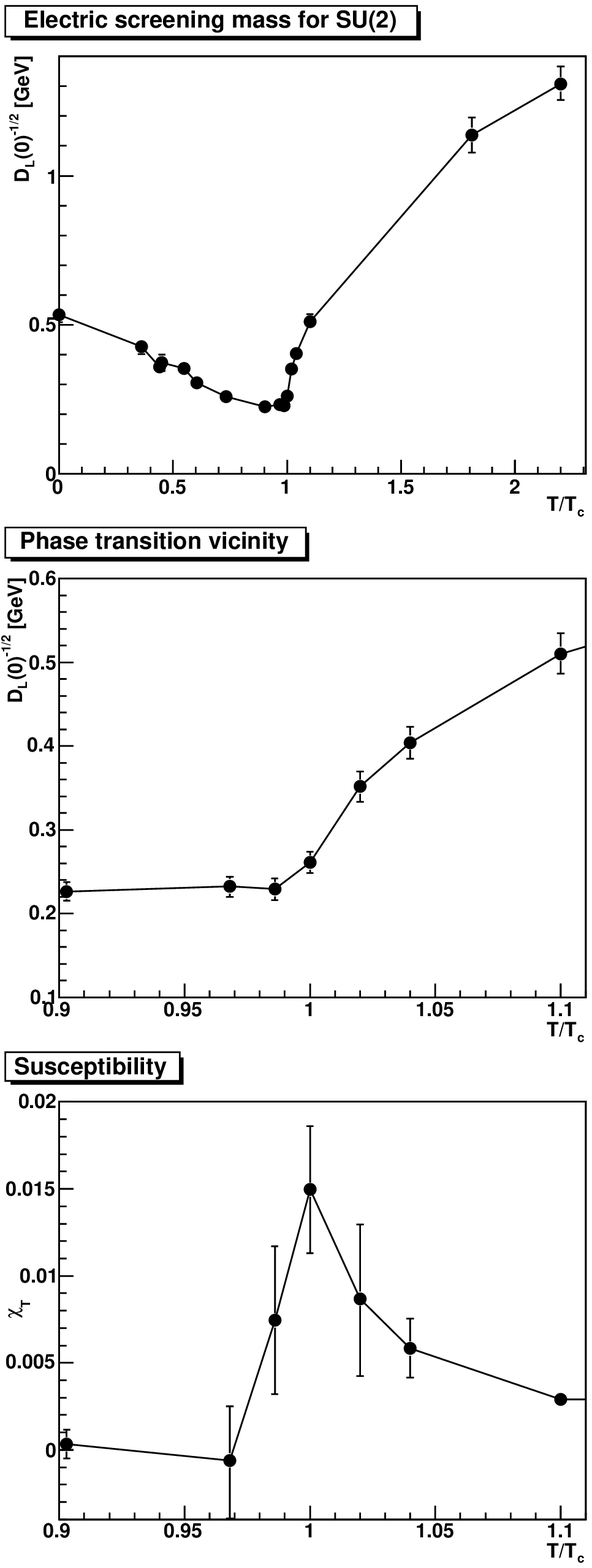}\hfill 
\includegraphics[width=0.4\linewidth]{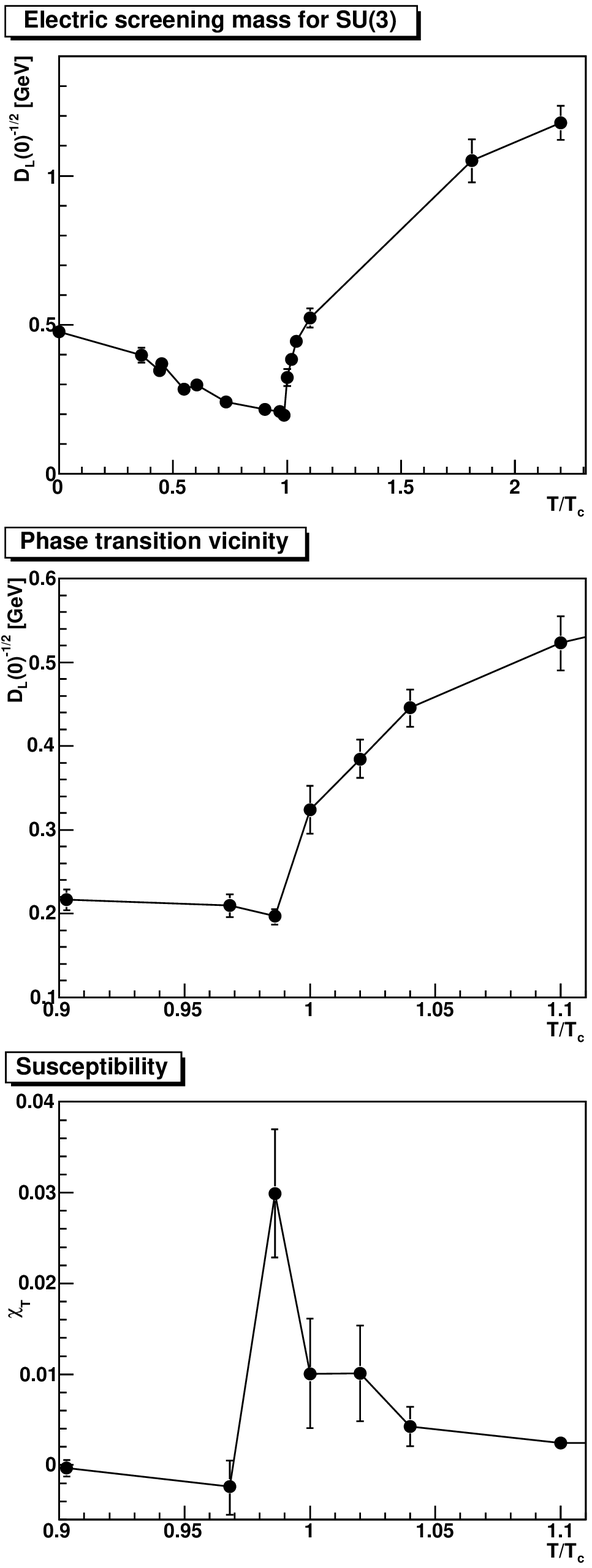}
\caption{\label{lat-em}The electric screening mass as a function of temperature 
(top panel), and zoomed in the phase transition region (middle) panel. The 
corresponding susceptibility $\chi_T=\pd D_L(0)^{-1/2}/\pd T$ is shown in the 
bottom panel. On the left results for SU(2) and on the right results for SU(3) 
are shown. Note the respective different range for SU(2) and SU(3) for the 
susceptibility. Lines are drawn to guide the eye.}
\end{figure*}

\section{The gluon propagator from lattice calculations}\label{sec:lattice}

\subsection{Simulation details}

To determine the gluon propagator between zero and roughly two times the critical 
temperature ($T_c=277$ MeV for SU(2) and $T_c=303$ MeV for SU(3)), lattice gauge theory implementing the Wilson action is used, details 
can be found in \cite{Cucchieri:2007ta,Cucchieri:2006tf}. Details of the simulation 
parameters can be found in appendix \ref{app:simul}, in particular its table \ref{conf}.
There also systematic errors will be discussed. This investigation extends previous 
works in the same temperature range \cite{Cucchieri:2007ta,Karsch:1994xh}, 
and at higher temperatures \cite{Cucchieri}.

At finite temperature, the Landau gauge gluon propagator $D_\mn$ is described
by two independent dressing functions, $D_{T}$ and $D_{L}$,
\bea
D_\mn^{ab}(p)&=&P_\mn^T(p)D_{T}^{ab}(\op^2,\vec p^2)+P_\mn^L D_L^{ab}(\op^2,\vec p^2)\nn\\
P_\mn^T(p) &=& (1-\delta_{\mu0})(1-\delta_{\nu0})\left(\delta_\mn-\frac{p_\mu p_\nu}{\vec p^2}\right)\nn\\
P_\mn^L(p) &=& P_\mn(p)-P_\mn^T(p)\nn\,,
\eea
where $P_\mn^T$ and $P_\mn^L$ are projectors transverse and longitudinal to the
heat bath. The color-dependency of the propagator has been evaluated in lattice gauge theory, and has 
always been found to be consistent with a color-diagonal propagator \cite{Cucchieri:2006tf}. 
Thus there remain two scalar functions $D_T=Z_T/p^2$ (transverse to the heat bath's four-velocity, 
chromomagnetic) and $D_L=Z_L/p^2$ 
(parallel to the heat bath's four-velocity, chromoelectric), with $Z_T=Z_L$ at zero temperature, 
for the description of the gluon. 
Here $p$ denotes always the four-momentum with $p^2=\op^2+\vec p^2$, which 
coincides with the three-momentum for the soft modes $\op=0$. Only this soft mode will be discussed in 
detail here. The hard modes are discussed separately in appendix \ref{app:hard}, 
and found to be very well approximated by $D(\op^2,\vec p^2)=D(0,\op^2+\vec p^2)$.

The gluon propagator, as the gluon itself, is only well-defined in a fixed gauge. 
For the purpose of calculating gauge-invariant quantities the choice of gauge is irrelevant, 
but since gauge-dependent intermediate results will be transferred from one method 
to another here, a non-perturbatively unambiguous definition of the gauge seems necessary. 
We choose here the minimal or average-$B$ Landau gauge \cite{Cucchieri:gf,Maas:2009se}. 
For the discussions of other choices see 
\cite{Maas:2009se,Maas:2008ri,Bogolubsky:2009qb,Silva:2004bv}. However, the gluon 
propagator at the volumes employed here is depending only marginally on this choice 
\cite{Maas:2009se,Bogolubsky:2009qb}. In particular, the effect is likely only of 
the same order as other systematic effects of the present calculations. Therefore, 
it can also be used for similar other calculations, independent of the particular 
choice between all non-perturbative realizations of the Landau gauge. The methods 
used to fix this gauge at finite temperature can be found in \cite{Cucchieri:2007ta} 
for SU(2), and the extension used to address SU(3) in \cite{Maas:2007af}. In 
\cite{Cucchieri:2007ta,Cucchieri:2006tf} also the methods used to determine the 
gluon propagator are presented. Note that the method to determine the value 
of the gluon fields employed in general and in particular in \cite{Cucchieri:2007ta} 
is only applicable to configurations with positive real part of the Polyakov loop 
\cite{Karsch:1994xh}. Therefore, only such configurations have to be included in 
the Monte-Carlo average which satisfy this condition. This is performed by filtering 
at all temperatures for this condition.

In addition to the gluon we determine the ghost propagator $D_G$, related to its dressing function by
\beq
D_G(p)= \frac{G(p)}{p^2} 
\eeq
and found previously to be color-diagonal \cite{Cucchieri:2006tf}. In contrast to 
the gluon propagator, the ghost is expected to depend significantly on the gauge 
choice for momenta $p < 1$ GeV \cite{Maas:2009se,Maas:2008ri,Bogolubsky:2009qb}. 
Here, the only reason for its investigation is to study the prediction of functional 
methods \cite{Cucchieri:2007ta} that the gauge-fixing dynamics, which is essentially given by the ghost 
propagator, is insensitive to the phase transition. This is indeed the case as
already indicated by the results of Ref.~\cite{Cucchieri:2007ta} and further confirmed
by our results for the ghost discussed in appendix \ref{app:ghost}. 

All results presented are renormalized such that at 2 GeV the dressing functions in all 
cases equal 1, for convenience. The renormalization factor is determined by the average 
value of the propagators at the closest momentum values above and below 2 GeV in the 
$x$-$y$-plane. This procedure entails that the renormalization constants are 
temperature-dependent. They vary by about 10\% for the transverse gluon and the ghost, 
but by up to 50\% for the longitudinal gluon in the investigated temperature and 
discretization range. 

\subsection{Results}\label{res:lattice}

The finite-temperature results for the transverse part $D_T$ of the 
gluon propagator are shown in figure \ref{lat-gpt}. In the left panel we display 
the results for SU(2) and in the right panel corresponding results for SU(3). The
temperature behaviour of each momentum mode measured can be read off the 3d-plot
at the bottom of each panel. In general we observe a relatively smooth variation
of all large momentum modes of the propagator across the phase transition in agreement 
with previous expectations \cite{Cucchieri:2007ta}. These modes behave similar for SU(2)
and SU(3). Only in the infrared some more drastic variations can be seen. These appear 
to be systematic close to the phase transition temperature and are somewhat more 
pronounced for SU(3). Although these variations are clearly not statistical in nature,
it is not clear whether they are genuine. In particular, they may
be volume effects, since increased long-wavelength thermal fluctuations close to 
the phase transition are not faithfully represented by the volumes presently employed.
This possibility is supported by the fact that the effect is stronger in the case of 
SU(3) than of SU(2) where the employed volumes are smaller. In any case, more refined 
and systematic studies are necessary to explore the nature of these variations. 

In general we observe that the chromomagnetic, transverse sector of Yang-Mills theory
shows no pronounced reaction to the phase transition. The only essential effect is that with 
increasing temperature the propagator becomes stronger infrared suppressed and develops
even a maximum at $p \approx 0.5$ GeV for the largest temperatures investigated here. 
This is indeed expected given that the propagator in the infinite-temperature limit of the 
dimensionally reduced theory in this gauge exhibits 
a clear maximum \cite{Cucchieri:2003di}.

The situation is rather different for the chromoelectric, longitudinal part $D_L$ of the
gluon propagator, shown in figure \ref{lat-gpl}. From the plots it is immediately visible 
that the longitudinal gluon reacts strongly to the phase transition, in accordance with 
previous observations \cite{Cucchieri:2007ta}. For temperatures below the phase transition
we observe a dramatic increase of the infrared part of the propagator close to the
phase transition which leads to an even more dramatic decrease shortly above $T_c$. Also
this dramatic variation seems to be more pronounced in the SU(3) data. This behaviour
is most easily visible from the electric screening mass 
\be
m_L = D_L(p \rightarrow 0)^{-1/2}\nn,
\ee 
and its associated susceptibility
\be
\chi_T=\frac{\pd m_L}{\pd T}\nn,
\ee
\no which are shown in figure \ref{lat-em}. This mass drops when moving towards 
the phase transition, reaches a minimum at the phase transition, and then quickly 
increases. In fact, the sensitivity is sufficiently strong that the gauge-invariant 
critical temperature can be determined independently from the chromoelectric 
screening mass just by extracting it from the plot. Indeed, this temperature coincides with the 
independently determined one using the Polyakov loop as an order parameter 
\cite{Fingberg:1992ju,Lucini:2003zr} within the resolution of the temperature grid 
employed here. 

In the middle panel of figure \ref{lat-em} we zoom into the temperature region around
the phase transition. From the available data, it appears that the screening mass in 
the SU(3) case is changing more rapidly across $T_c$ than for the SU(2) case. 
It may be that this more rapid change signifies the first-order nature of the SU(3) 
transition, whereas in the SU(2) case the behavior is smoother as may be expected for a 
second-order phase transition, despite the larger physical volumes. This interpretation 
may also be supported by the much larger susceptibility in the SU(3) case, shown 
in the lower panel of figure \ref{lat-em}. However, at present this interpretation 
may only indicate a possible scenario and certainly has to be checked in a more 
detailed analysis \cite{Maas:wip}. 
In addition, a careful volume study is needed to check for the presence of long range
correlations in the form of a vanishing electric screening mass of the SU(2) gluon at $T_c$.  

Together, our results for the chromoelectric and chromomagnetic part of the gluon 
propagator indicate that the dominant response to the phase transition occurs in the 
chromoelectric sector. This finding can be related to the behaviour of certain 
dimension-two condensates \cite{Chernodub:2008kf,Dudal:2009tq}. These condensates are 
effectively given by the difference between the traces of the transverse and the 
longitudinal gluon propagator \cite{Maas:2008ri}. Thus the asymmetry
observed in these condensates at finite temperatures \cite{Chernodub:2008kf} can be attributed 
essentially to the behavior of the longitudinal gluon propagator alone.

These comments complete our investigation of the temperature dependent gluon propagator. 
In the next section we will see that the gluon data are an important input into the
Dyson-Schwinger equation for the quark propagator and therefore serve as a basis for
our investigation of the order parameters for the chiral and deconfinement
transition of quenched QCD.

\section{Order parameters extracted from the quark propagator \label{sec:dse}}

\subsection{The conventional and the dual quark condensate}

The temperature dependent general expression for the inverse quark propagator $S^{-1}$
is given by
\bea  \label{quark}
S^{-1}(\vp,\op) &=& i \gamma_4\, \op C(\vp,\op) 
\nn\\[0.1cm]&&+ i \gamma_i \, p_i A(\vp,\op)+ B(\vp,\op) \,,
\eea
with vector and scalar quark dressing functions $C,A,B$. A further tensor component 
proportional to $\sigma_{\mu \nu}$ is possible in principle but is negligible in the 
presence of a pure vectorial quark-gluon vertex 
\cite{Roberts:2000aa}, as used here. The momentum arguments are given in
terms of the three momenta $\vp$ and generalised Matsubara frequencies $\op$
\beq
\op(n_t,\varphi) = (2\pi T)(n_t+\varphi/2\pi). 
\eeq
These correspond to $U(1)$ valued boundary conditions for the quark fields, i.e.
$\psi(1/T,\vec{x}) = e^{i \varphi} \psi(0,\vec{x})$. The angle $\varphi$ varies 
between $\varphi \in [0,2\pi[$, with $\varphi=0$ for periodic and $\varphi=\pi$
for the usual, physical antiperiodic boundary conditions for the fermionic quarks.

Given the momentum behaviour of the non-perturbative dressing functions $C,A,B$ one 
can extract a $\varphi$-dependent quark condensate from the propagator according to
\begin{align} \label{trace}
 \langle\bar{\psi}\psi \rangle_{\varphi} = 
 Z_2\, N_c\, T\sum_{\omega_p(\varphi)}\int\frac{d^3p}{(2\pi)^3}\,
 \textrm{tr}_D\,S(\vec{p},\omega_p(\varphi))\,.
\end{align}
The conventional quark condensate is obtained for the special case $\varphi=\pi$
and multiplication of this expression with $Z_m$. In the limit of
vanishing bare quark masses it is an order parameter for the chiral phase transition.

The corresponding dual observable, the dual quark condensate or dressed Polyakov loop,
is obtained by a Fourier-transform of the $\varphi$-dependent condensate with respect
to the winding number $n$,
\beq \label{dual}
\Sigma_n = \int_0^{2\pi} \, \frac{d \varphi}{2\pi} \, e^{-i\varphi n}\, 
\langle \overline{\psi} \psi \rangle_\varphi
\eeq
and specialising to the case $n=1$. The dressed Polyakov loop $\Sigma_1$ is sensitive to
the breaking and restoration of center symmetry \cite{Jena1,dualcond1} and therefore
serves as an order parameter for the deconfinement transition.

These two order parameters, \Eq{trace} with $\varphi=\pi$ and \Eq{dual}, can be extracted from the dressed,
temperature dependent quark propagator calculated from functional methods as detailed
in Refs.~\cite{Fischer:2009wc,Fischer:2009gk,Braun:2009gm}. In the following we will
update the calculation presented in Ref.~\cite{Fischer:2009gk} and in addition also extend
 the studies to the case of the gauge group SU(3). We will investigate 
whether and how the better temperature resolution of the lattice data for the gluon
propagator discussed in the previous section leads to improved results for the order
parameters. We also compare the results obtained for gauge group SU(2) and SU(3).

\subsection{Truncation scheme for the Dyson-Schwinger equations for the quark 
propagator \label{trunc}}

\begin{figure*}[t]
\includegraphics[width=0.9\columnwidth]{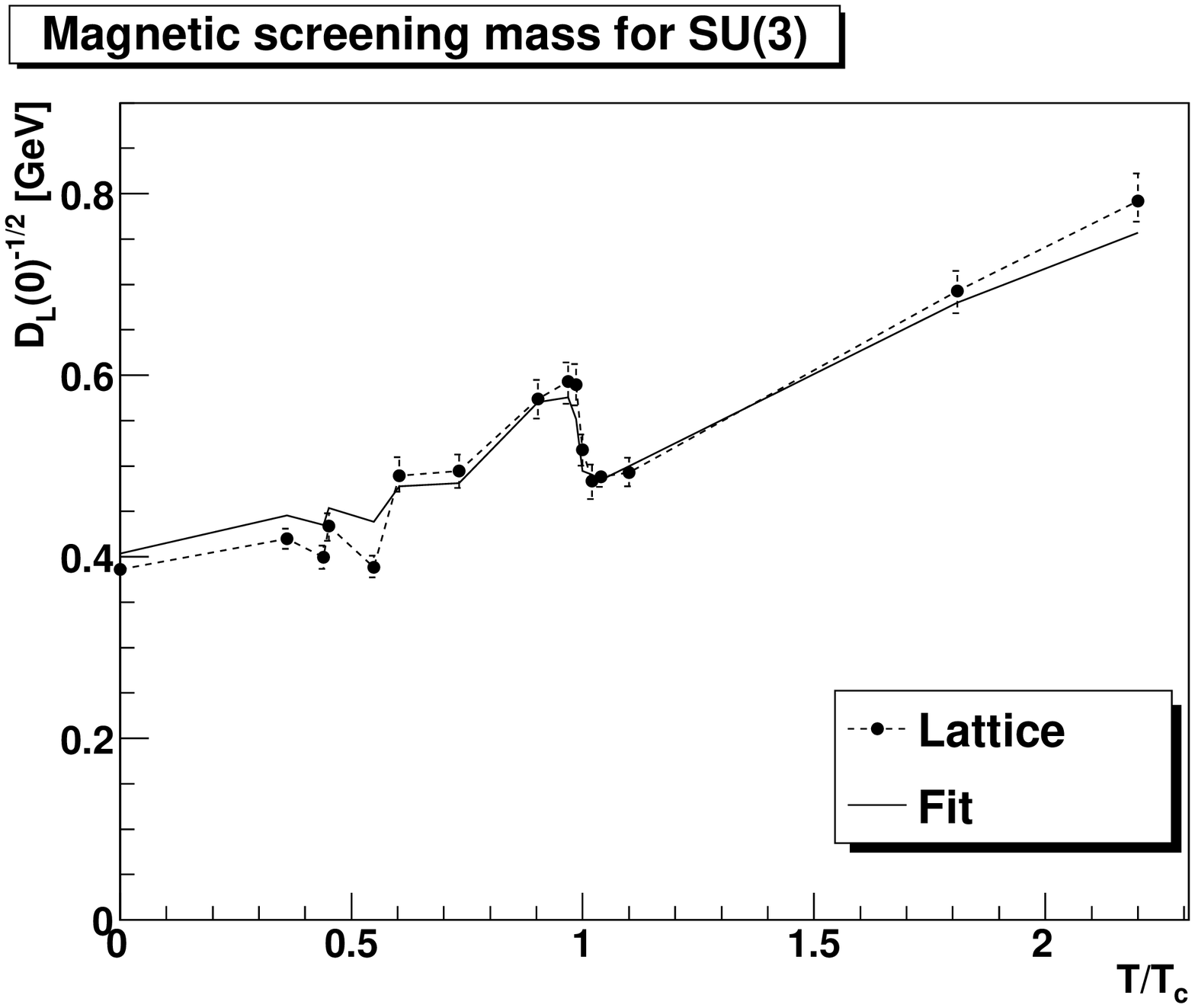}\hfill
\includegraphics[width=0.9\columnwidth]{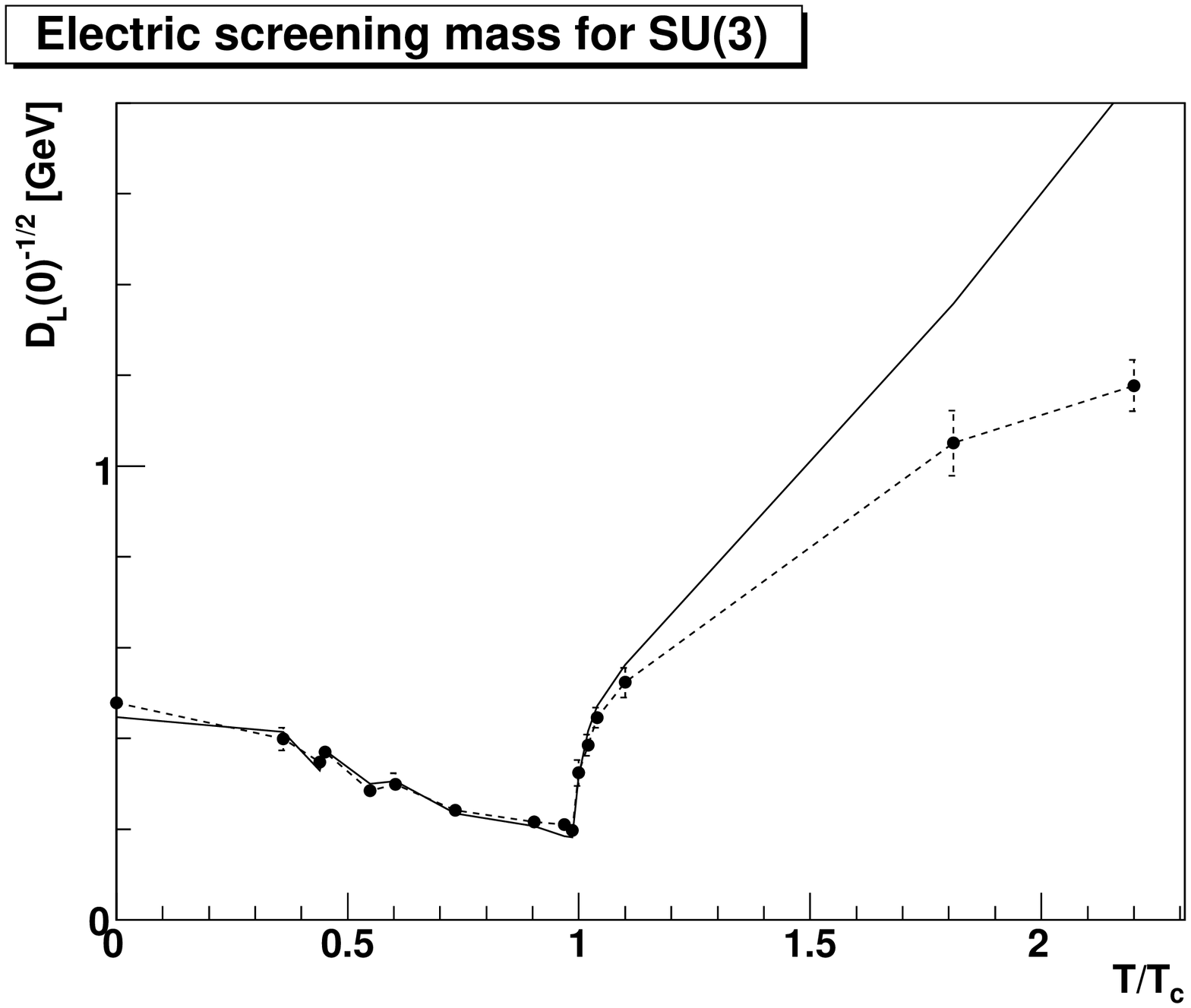}
\caption{
Magnetic (left) and electric (right) temperature dependent scale parameters
in our fits for the lattice gluon propagator compared with the lattice
results for the magnetic and electric screening masses of the gluons.\vspace*{-2mm}}
\label{res:fits}
\end{figure*}
The Dyson-Schwinger equation for the quark propagator at
finite temperature $T$ is given by 
\beqa \label{DSE}
S^{-1}(\vp,\op) &=& Z_2 \, S^{-1}_0(\vp,\op) 
-  C_F\, \frac{Z_2 \widetilde{Z}_1}{\widetilde{Z}_3}\, g^2 T \intk \nn\\[0.1cm]
&&\times\gamma_{\mu}\, S(\vk,\ok) \,\Gamma_\nu(\vk,\ok,\vp,\op) \nn\\[0.1cm]
&&\times D_{\mu \nu}(\vp-\vk,\op-\ok) \,. \label{quark_t}
\eeqa
Here $D_{\mu \nu}$ denotes the (transverse) gluon propagator in Landau gauge and 
we have introduced a reduced quark-gluon vertex $\Gamma_\nu$, by defining
$\Gamma^{full}_{\nu,i} = i g \frac{\lambda_i}{2} \Gamma_\nu$. The 
bare quark propagator is given by 
$S^{-1}_0(p) = i \gamma \cdot p + Z_m m(\mu^2)$, where $m(\mu^2)$ is the 
renormalized current quark mass.  The wave function and quark mass renormalization 
factors, $Z_2$ and $Z_m$, are determined in the renormalization process. The ghost 
renormalization factor $\widetilde{Z}_3$ is canceled by a corresponding factor in 
our model for the quark-gluon vertex discussed below. Furthermore we used 
$\tilde{Z}_1=1$ for the renormalization factor of the Landau gauge ghost-gluon vertex.
The quark dressing functions $A(\vp,\op),B(\vp,\op)$ and $C(\vp,\op)$ can be extracted 
from Eq.~(\ref{DSE}) by suitable projections in Dirac-space. 

In \Eq{DSE} the Casimir factor $C_F = (N_c^2-1)/(2N_c)$ stems from the color trace.
For $N_c=2$ these equations have been solved numerically in Refs.~\cite{Fischer:2009wc,Fischer:2009gk}
in a truncation scheme which used lattice results from Ref.~\cite{Cucchieri:2007ta} 
for the temperature dependent gluon propagator as input. As discussed in the previous 
section we now have updated and refined results for the SU(2) and SU(3) gluon propagator
at our disposal. These will be used in the following to provide for improved results
for the quark propagator and the associated conventional and dual condensates.

In the Dyson-Schwinger equation (\ref{DSE}) for the quark propagator we need to
evaluate the gluon propagator at values of momentum which are not identical with the
ones extracted on the lattice. In particular we need to evaluate the gluon at momenta
larger than available lattice momenta. As it turns out, it is possible to
use our knowledge on the analytical form of the gluon dressing functions $Z_{T,L}$ at 
zero temperature \cite{Alkofer:2003jk} to devise a fit function which is 
capable to nicely interpolate and extrapolate the lattice data also at finite temperature.
These fits for the transverse and longitudinal dressing functions $Z_{T,L}$ of the 
gluon propagator are given by
\beqa \label{gluefit}
Z_{T,L}(\vq,\oq,T) &=& \frac{q^2 \Lambda^2}{(q^2+\Lambda^2)^2} \,
\Bigg{\{}\left(\frac{c}{q^2+ \Lambda^2 a_{T,L}(T)}\right)^{b_{T,L}(T)} \nn\\[0.1cm]
&&+\frac{q^2}{\Lambda^2}\left(\frac{\beta_0 \alpha(\mu)\ln[q^2/\Lambda^2+1]}{4\pi}\right)^\gamma\Bigg{\}}\,.
\eeqa
Here we find a temperature independent scale parameter $\Lambda = 1.4$ GeV and 
the coefficient $c=11.5 \,\mbox{GeV}^2$. Furthermore $\beta_0 = 11N_c/3$ 
and $\gamma=-13/22$ in the quenched theory and we renormalize at 
$\alpha(\mu)=0.3$.

The fit function \Eq{gluefit} generalises the one used in 
Refs.~\cite{Fischer:2009wc,Fischer:2009gk}, where the temperature 
dependent exponent $b_{T,L}(T)$ has been kept fixed, i.e. $b_{T,L}(T)=2$.
Since we now have more accurate lattice gluon data at our disposal we 
found it useful to relax this condition and thus provide an even better 
representation of the lattice data. One may speculate whether an irrational 
exponent $b_{T,L}(T)$ with the corresponding temperature dependent cut in 
the $p^2$-plane signals quantitative changes also in the analytic structure
of the gluon, as suggested in \cite{Maas:2005hs,Maas:2004se}. A detailed 
investigation of this aspect is relegated to further 
systematic studies \cite{Maas:wip}.

In general our fits are optimised in particular in the mid-momentum
regime, which is most important later on when we use them as input into the quark-DSE.
The details for the fit parameters as well as plots of the fits for selected 
temperatures are relegated to appendix \ref{app:fit}. Here we only show the 
resulting electric and magnetic screening masses of the gluon as extracted from 
the infrared behaviour of our fit functions. These are compared to the 
corresponding lattice results in fig.~\ref{res:fits}. The fit quality is very satisfactory, except for the electric screening mass at the largest temperatures.
Here the lattice results suffer mostly from systematic artifacts due to the 
restricted number of points in the time direction \cite{Cucchieri:2007ta}. This may be reflected in
the mismatch of the screening masses with the ones extracted from our fit. 
Indeed, one may even argue that the electric screening masses from the fits are
more accurate in the large temperature regime since they nicely reproduce
the expected proportionality of the screening mass with temperature,
$m_{L} \sim T$, known from hard thermal loop results.
We also note that at the two highest temperatures available the fit function describes the low and mid momentum
behavior of the electric gluon propagator very precise. On the other hand the fit function \Eq{gluefit}
is not capable to describe the qualitative mid-momentum dependence of the magnetic propagator
 in this temperature range, as can be seen in fig.~\ref{res:fitscompared} in the appendix. 
We checked that this momentum behavior can be described precisely using in addition a
momentum dependent screening term in the fit function.
 Anyway, in the important region around the critical temperature both fits
work perfectly well and represent therefore a trustable input for the 
Dyson-Schwinger equation of the quark propagator.

Note that as a significant difference to the fits used in 
Refs.~\cite{Fischer:2009wc,Fischer:2009gk} it turns out that the transition
of the electrical screening mass from its decreasing behaviour below the
critical temperature $T_c$ to the increase above $T_c$ is much sharper than
the one extracted in Refs.~\cite{Fischer:2009wc,Fischer:2009gk}. This sharp
change around $T_c$ was not resolved by the then available lattice 
data of Ref.~\cite{Cucchieri:2007ta}. As a consequence of the much improved 
temperature resolution available now we will see that the corresponding 
deconfinement transition extracted from the quark propagator is also much 
more pronounced than the one seen in \cite{Fischer:2009wc,Fischer:2009gk}.
This will be discussed in more detail in the next section.

The remaining piece to be specified in the quark-DSE is the dressed quark-gluon
vertex. Similar to Refs.~\cite{Fischer:2009wc,Fischer:2009gk} we employ the 
following temperature dependent model
\beqa \label{vertexfit}
\Gamma_\nu(q,k,p) &=& \widetilde{Z}_{3}\left(\delta_{4 \nu} \gamma_4 
\frac{C(k)+C(p)}{2}
+  \delta_{j \nu} \gamma_j 
\frac{A(k)+A(p)}{2}
\right)\nn\\[0.1cm]
&&\times\Bigg{(} 							
\frac{d_1}{d_2+q^2} 			
 + \frac{q^2}{\Lambda^2+q^2}\nn\\[0.1cm]
&&\times\bigg(\frac{\beta_0 \alpha(\mu)\ln[q^2/\Lambda^2+1]}{4\pi}\bigg)^{2\delta}\Bigg) \,,
\eeqa 
where $q=(\vq,\oq)$ denotes the gluon momentum and $p=(\vp,\op)$, $k=(\vk,\ok)$ the
quark and antiquark momenta, respectively. Furthermore $2\delta=-18/44$ is the anomalous dimension 
of the vertex. Both together, the gluon dressing function and the quark-gluon vertex 
behave like the running coupling at large momenta; this is a necessary boundary 
condition for any model interaction in the quark DSE. The dependence of the vertex 
on the quark dressing functions $A$ and $C$ is motivated by the Slavnov-Taylor 
identity for the vertex; it represents the first term of a generalization of the 
Ball-Chiu vertex \cite{Ball:1980ay} to finite temperatures. The remaining fit 
function is purely phenomenological, see e.g. \cite{Fischer:2008wy} where an 
elaborate version of such an ansatz has been used to describe meson observables. 
We use $d_2=0.5 \,\mbox{GeV}^2$ for both gauge groups, but $d_1 = 7.6 \,\mbox{GeV}^2$ 
for SU(2) and $d_1=4.6 \,\mbox{GeV}^2$ for SU(3).
The change in parameter $d_1$ from SU(2) to SU(3) is again motivated by the Slavnov-Taylor 
identity. At high temperatures it is expected that it reduces 
to the QED Ward-Takahashi identity multiplied with the non-perturbative ghost dressing 
function. A comparison of SU(2) and SU(3) ghost dressing functions in the infrared 
calculated on the lattice shows that for small momenta, $G(p)$ of SU(3) is reduced by 
roughly half compared to SU(2), see fig.~\ref{lat-ghp}. Even though the quantitative 
values for the ghost dressing functions at low momenta from the lattice might
contain considerable uncertainties we assume the ratio of SU(2) to SU(3) to be reliable. 
In addition we also checked that a moderate variation of these parameters does not shift 
the critical temperatures of both, 
the chiral and the deconfinement transition.

Finally we wish to repeat a word of caution as concerns the chiral limit in our
approximation scheme \cite{Fischer:2009gk}. A prominent feature 
of the quenched theory not reproduced by our framework is the appearance of 
quenched chiral logarithms in the chiral condensate. These are well-known
to be generated by $\eta'$ hairpin diagrams, which are not represented by
our vertex ansatz. For the present investigation we believe this is
more an advantage than a drawback. Quenched chiral logarithms are most 
notable in the chiral limit, where they lead to a singularity in the chiral 
condensate. Since we do not encounter this singularity we are in a position
to investigate both, the ordinary condensate and the dressed Polyakov loop 
also in the chiral limit.

The numerical methods needed to solve the quark-DSE have been explained in detail
in Ref.~\cite{Fischer:2009gk}. Having specified all necessary input we now proceed
to present our results in the next subsection.

\subsection{Numerical results \label{sec:num}}
\begin{figure*}[t]
\includegraphics[width=0.9\columnwidth]{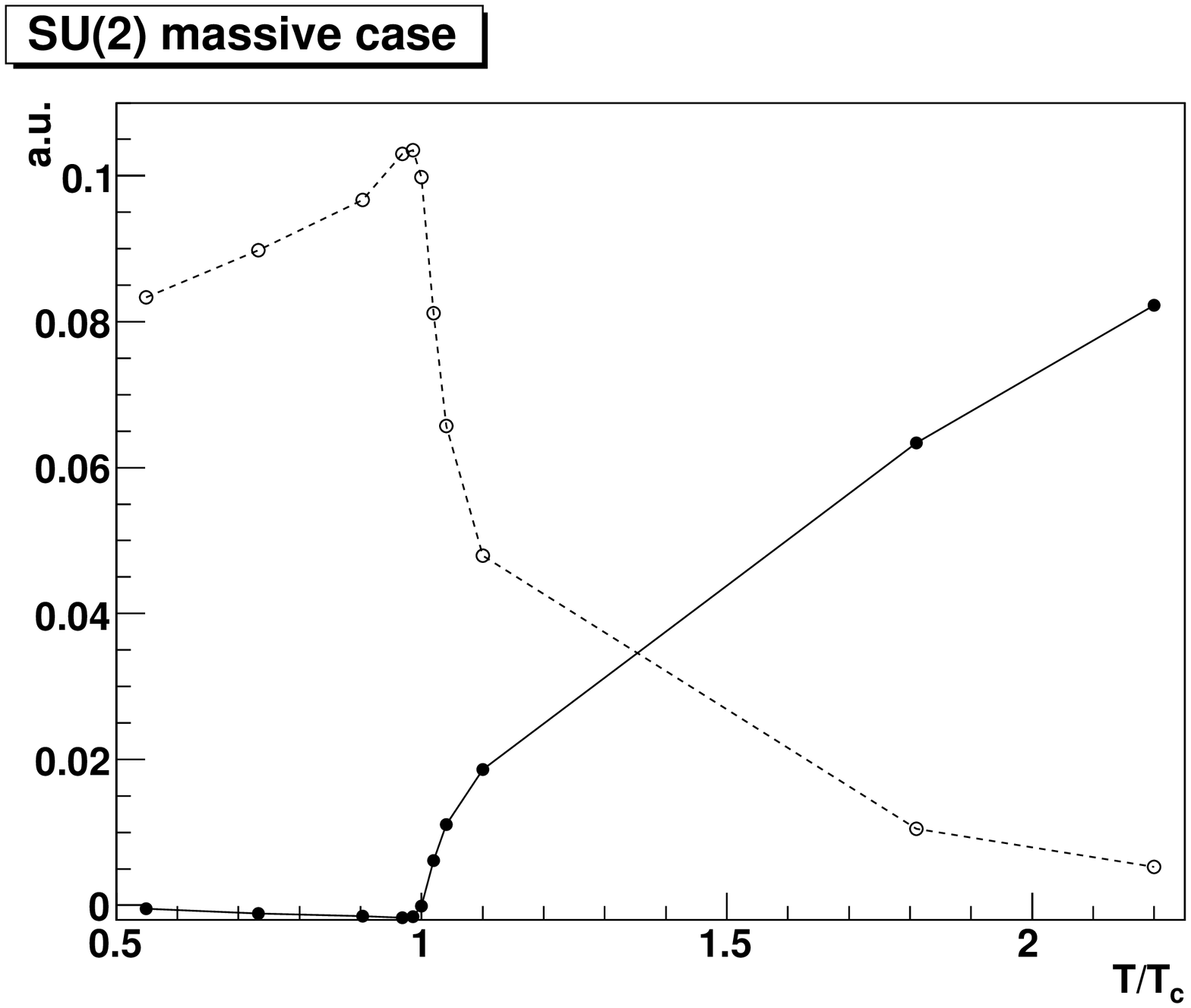}\hfill
\includegraphics[width=0.9\columnwidth]{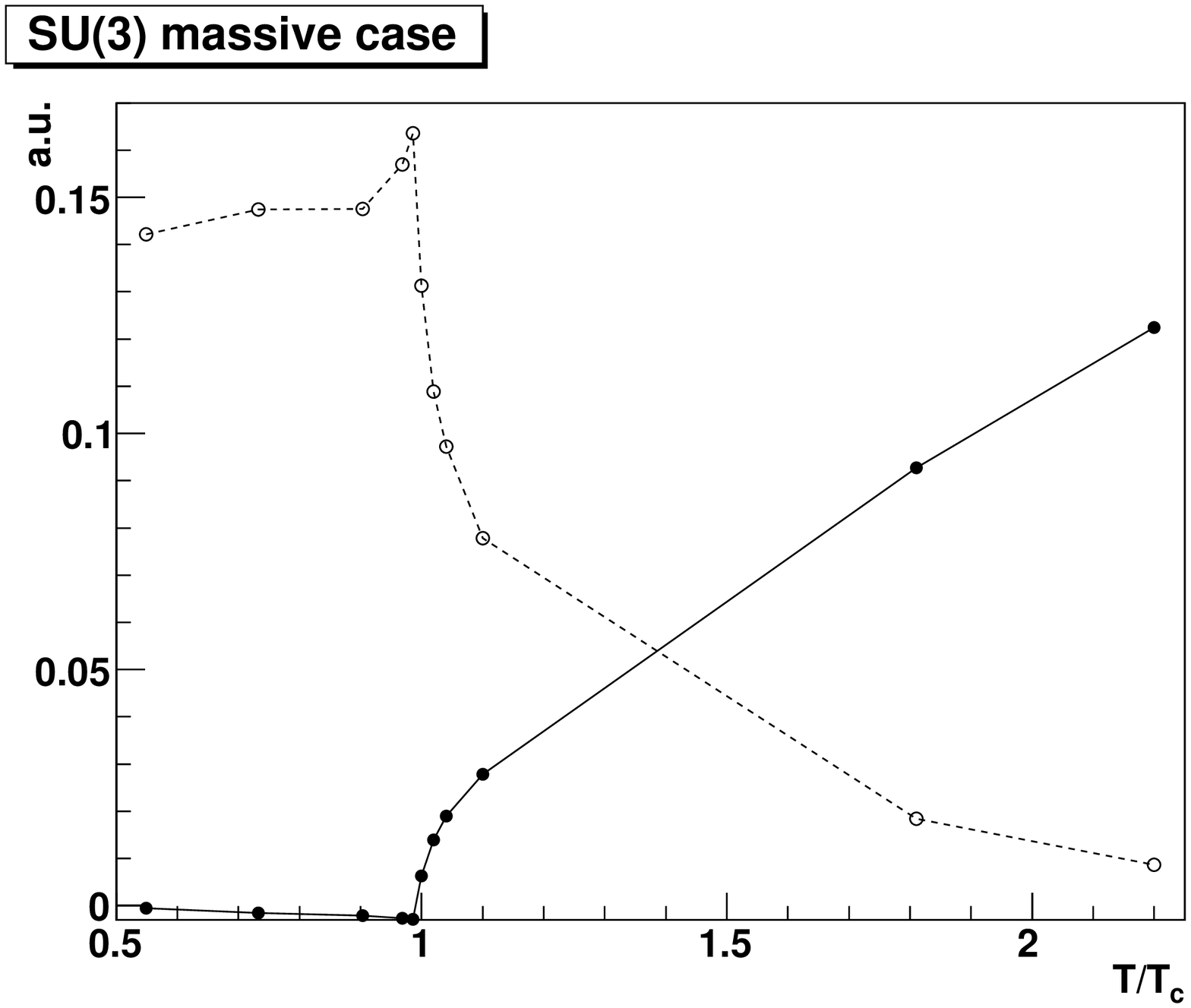}\vspace*{5mm}
\includegraphics[width=0.9\columnwidth]{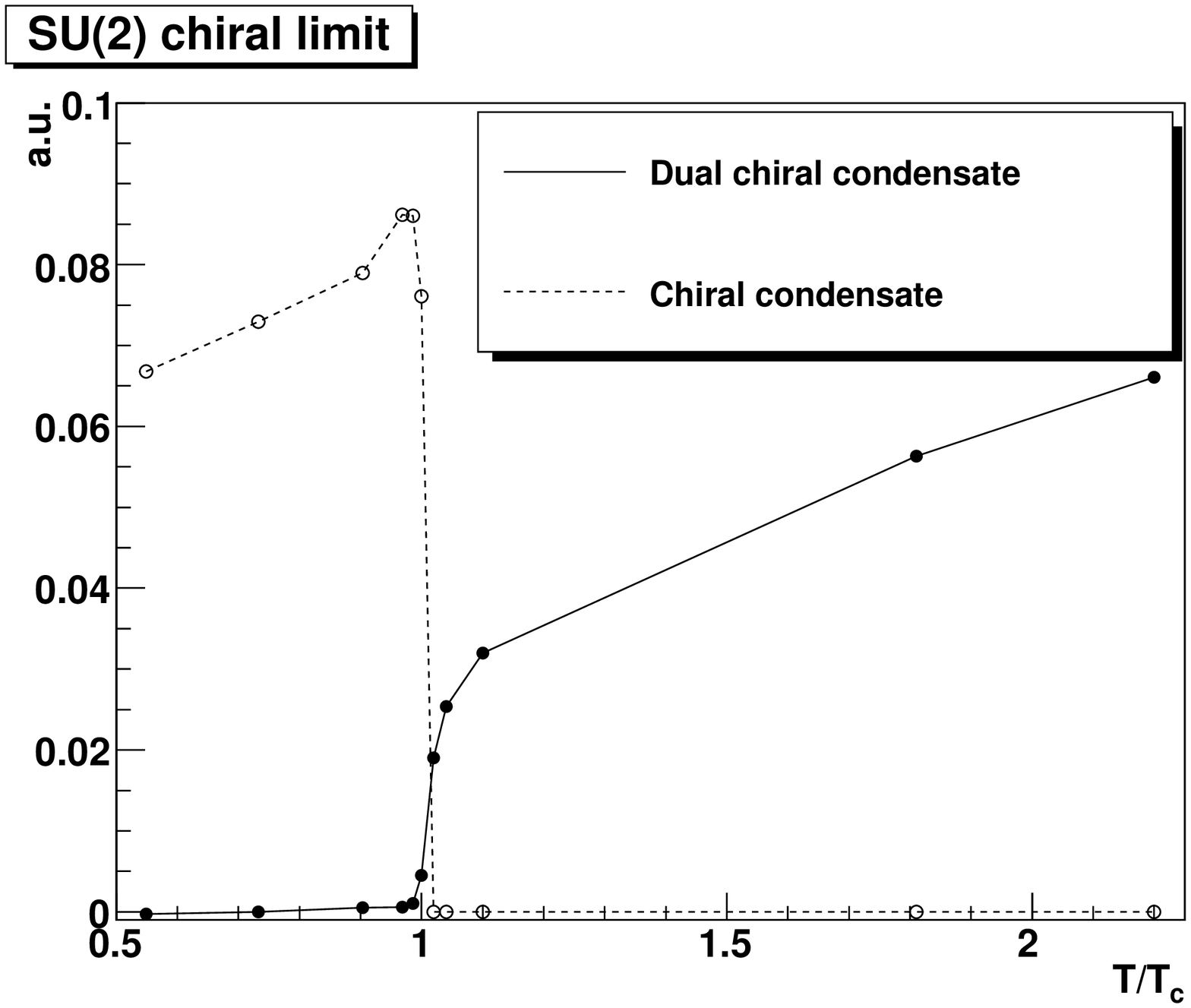}\hfill
\includegraphics[width=0.9\columnwidth]{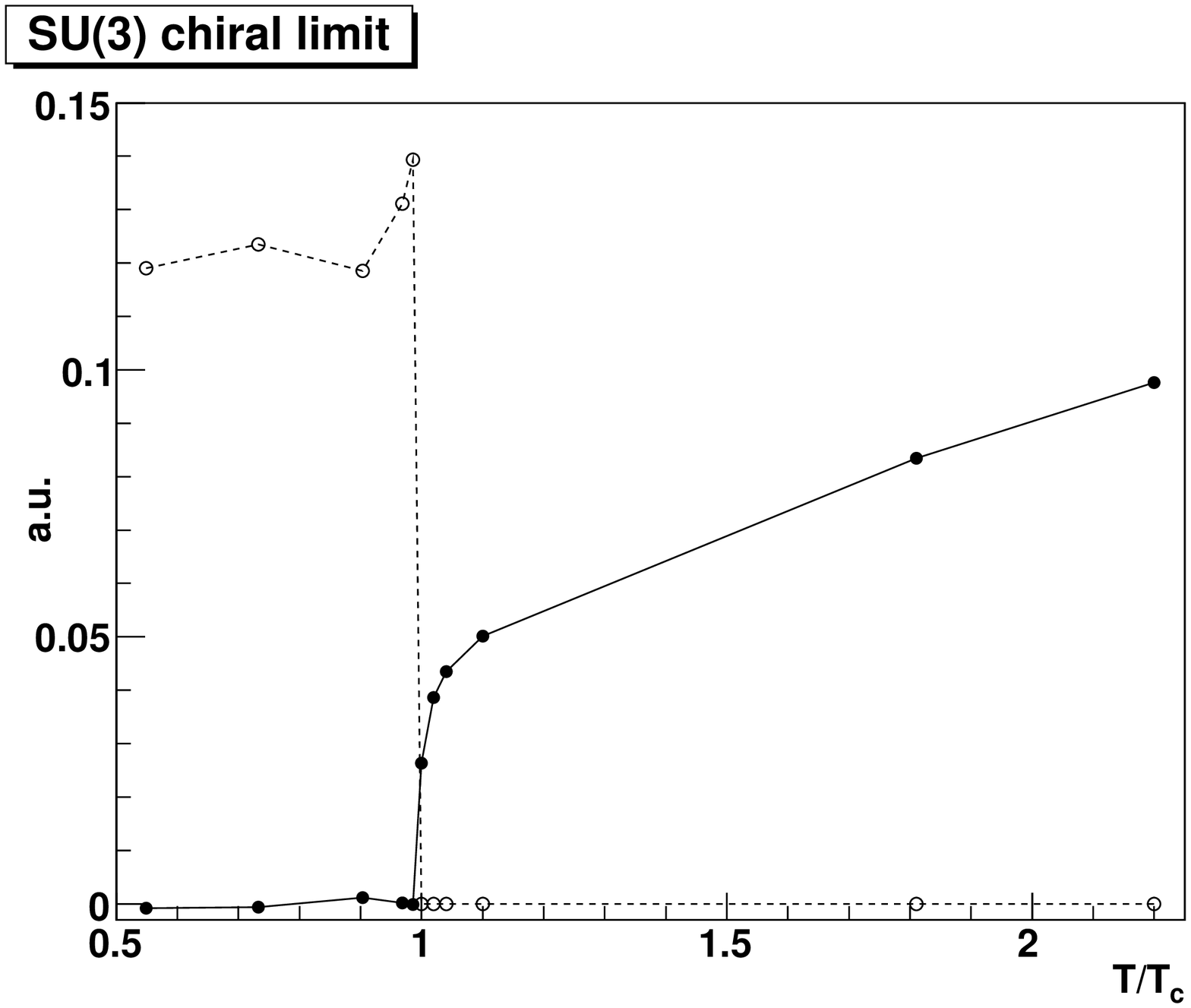}
\caption{
The quark condensate $\langle\bar{\psi}\psi \rangle_{\pi}$ and the dressed 
Polyakov loop $\Sigma_1$ as a function of 
temperature for SU(2) (left panel) and SU(3) (right panel) Yang-Mills theory.
Shown are results in arbitrary units (a.u.) for a massive (strange-)quark (upper panel) with 
$m \approx 80$ MeV and in the chiral limit (lower panel).}
\label{res:cond}
\end{figure*}
Our numerical results for the ordinary quark condensate, \Eq{trace} with 
$\varphi=\pi$, and the dressed Polyakov loop, \Eq{dual}, are shown in 
Fig.~\ref{res:cond}. Let us first concentrate on the results for the
ordinary quark condensate. Both, for SU(2) and SU(3) we find chiral
transitions taking place on a very small temperature interval. This is
particularly clear in the chiral limit (lower panel) but also for quark 
masses as heavy as the strange quark mass (upper panel). This temperature
interval clearly coincides with the one identified in section \ref{res:lattice}
for the change of behaviour in the electric and magnetic screening masses.
Technically what happens in the quark-DSE is that below $T_c$ the electric,
longitudinal part of the gluon propagator increases dramatically and therefore
provides for extra interaction strength in the quark DSE. As a consequence,
we find increasing values for the quark condensate. At or below $T_c$ the electric
part of the gluon propagator reaches its maximum and then drops
sharply around and above $T_c$. In the quark-DSE this sudden loss of 
interaction strength is responsible for the dramatic decrease in the chiral 
condensate. It is satisfying to note that a similar behaviour has been
observed from calculations of the quenched condensate via Casher-Banks relations
on the lattice \cite{Buividovich:2008ip}. This agreement gives us
confidence that the temperature dependence of our truncation for the 
quark-gluon vertex, \Eq{vertexfit}, is at least qualitatively reliable
and leads to meaningful results.

In the chiral limit we clearly obtain a chiral phase
transition from the conventional quark condensate. Unfortunately, the temperature 
resolution of the lattice input is still not fine enough to unambiguously identify 
the order of the phase transition. One may speculate whether the SU(2) transition
is second or first order, whereas the SU(3) one seems to be first order. The behaviour
at finite quark masses may be compatible with a rapid cross-over for SU(2) and 
in the case of SU(3) even with a jump in the condensate signaling a remnant of a 
first order transition. Further investigations are necessary to clarify, whether 
the differences seen in Fig.~\ref{res:cond} between SU(2) and SU(3) are indeed 
significant. 

As concerns the dressed Polyakov-loop we clearly find a transition between
the center-symmetric low temperature phase and the center-broken phase at
transition temperatures very close to the ones encountered for the 
conventional quark condensate. Below $T_c$ the dressed Polyakov-loop
is almost constant and very small. For large quark masses close to the
transition temperature we even find small negative values of the Polyakov-loop.
We interpret these as artifacts introduced due to mass dependencies in the 
quark-gluon vertex that are not represented by our vertex ansatz. At temperatures 
$T_c \le T \le 1.1 T_c$ the Polyakov-loop rises sharply and
then less steeply for larger temperatures. Within the temperature range 
investigated we do not yet see a saturation of the dressed Polyakov-loop at 
large temperatures, although the results in the chiral limit may bear
some signals of such a behaviour. In general, the deconfinement transition
extracted from the dressed Polyakov-loop is as pronounced as the corresponding
signal in the electric and magnetic screening masses of the gluon propagator,
discussed in section \ref{res:lattice}. Similarly to the chiral transitions, 
also here the temperature resolution and the systematic quality of the input 
gluon propagator is not yet good enough to cleanly identify an order of the
deconfinement transition. 

For SU(2), the results shown here replace previous ones reported in 
Refs.~\cite{Fischer:2009wc,Fischer:2009gk}. As already discussed above,
the lattice data used as input in these works have been available only
for four different temperatures and had to be interpolated in between. 
This procedure generated a smooth behaviour of the gluon propagator 
around the critical temperature resulting also in a broad transition 
range as concerns the chiral condensate. Here, with our much better 
temperature resolution as concerns the lattice gluon, we are able to 
identify this behaviour as an artifact of the interpolation procedure.

\section{Summary}\label{sec:sum}

In this work we combined two non-perturbative methods, lattice calculations
and functional methods using Dyson-Schwinger equations, in a common framework to play on their 
individual strengths and to reduce the inherent problems of each approach.
In the lattice ab initio framework we determined the temperature behaviour
of the electric and magnetic parts of the Landau gauge gluon propagator
with fine enough temperature resolution to relate their behaviour to the
critical temperatures of the deconfinement transition of Yang-Mills theory.
For SU(2) and SU(3) we found a clear signal of the phase transition in the extracted 
electric screening mass and a less pronounced indication in the magnetic screening masses.
We also verified the earlier expectations
\cite{Cucchieri:2007ta} that the 'bump' in the electric dressing function
undergoes dramatic changes around $T_c$. These changes have been identified
as the source of equally dramatic changes in the ordinary quark condensate
as determined from the Dyson-Schwinger equation for the quark propagator:
the condensate keeps rising below $T_c$ only to decrease sharply at $T_c$.
This behaviour is in agreement with previous observations from lattice
calculations of the Casher-Banks relation \cite{Buividovich:2008ip}.
It also locates the mechanism for chiral symmetry restoration in Landau
gauge rather unambigously in the (ultra-)soft electric sector of the theory,
emphasizing its genuine non-perturbative nature.

As for the deconfinement transition we observe a clean transition from the
dressed Polyakov-loop as determined from the Dyson-Schwinger equation for 
the quark propagator. The corresponding transition temperatures, 
$T_c \approx 303$ MeV for SU(2) and $T_c \approx 277$ MeV for SU(3) agree 
with the ones extracted from our lattice calculations. For SU(2) our results 
for the dressed Polyakov-loop update and refine the ones of 
Refs.~\cite{Fischer:2009wc,Fischer:2009gk}, where only a much courser lattice 
input for the gluon propagator has been available.

Finally we wish to emphasize that the dressed Polyakov-loop examined in our
work is not the only gauge invariant order parameter for the deconfinement 
transition that can be calculated with functional methods. In Ref.~\cite{Braun:2007bx} 
the Polyakov-loop potential has been determined within the functional 
renormalization group and delivered the expected first and second order 
phase transition for SU(3) and SU(2), respectively.
 
In general we believe that all these results demonstrate that combined efforts of
the lattice and the functional framework provide a sound basis to determine
gauge invariant observables in gauge fixed Yang-Mills theory. They also provide a 
good starting point to generalise the approach towards the inclusion of unquenching
effects and the introduction of finite chemical potential.

\vspace{5mm}
\noindent
{\bf Acknowledgments}\\
We thank Lorenz von Smekal for helpful discussions.
C.\ F.\ and J.\ M.\ were supported by the Helmholtz Young Investigator 
Grant VH-NG-332, by the Helmholtz Alliance HA216-TUD/EMMI and the
Helmholtz International Center for FAIR within the LOEWE program of the 
State of Hesse.
A.\ M.\ was supported by the FWF under grant number M1099-N16. 
Part of the computing time was provided by the HPC center at the 
Karl-Franzens-University Graz. The ROOT framework \cite{Brun:1997pa} 
has been used in this project.
\begin{table*}[ht]
\caption{\label{conf}Data of the lattice gauge theory calculations as explained in the text.} 
\begin{center} 
\vspace{1mm}
\begin{tabular}{|c||c|c|c|c|c|c|c|c|}
\hline 
Group & $T/T_c$  & $T$ [MeV] & $N_t$ & $N_s$ & $\beta$ & $a$ [fm] & $V_s^{1/3}$ [fm] & conf. \cr
\hline\hline
SU(2) & 0        & 0         & 24    & 24    & 2.227   & 0.197    & 4.74                   & 356   \cr
SU(3) & 0        & 0         & 18    & 18    & 5.642   & 0.197    & 3.54                   & 142   \cr
SU(2) & 0        & 0         & 24    & 24    & 2.301   & 0.162    & 3.89                   & 189   \cr
SU(3) & 0        & 0         & 18    & 18    & 5.738   & 0.162    & 2.92                   & 82    \cr
\hline
SU(2) & 0.361    & 109       & 10    & 32    & 2.261   & 0.181    & 5.78                   & 169   \cr
SU(3) & 0.361    & 100       & 10    & 24    & 5.642   & 0.197    & 4.73                   & 94    \cr
\hline
SU(2) & 0.440    & 133       & 10    & 32    & 2.332   & 0.148    & 4.75                   & 194   \cr
SU(3) & 0.440    & 122       & 10    & 24    & 5.738   & 0.162    & 3.89                   & 73    \cr
\hline
SU(2) & 0.451    & 136       & 8     & 36    & 2.261   & 0.181    & 6.51                   & 115   \cr
SU(3) & 0.451    & 125       & 8     & 26    & 5.642   & 0.197    & 5.12                   & 73    \cr
\hline
SU(2) & 0.549    & 166       & 8     & 36    & 2.3315  & 0.149    & 5.35                   & 152   \cr
SU(3) & 0.549    & 152       & 8     & 26    & 5.738   & 0.162    & 5.12                   & 79    \cr
\hline
SU(2) & 0.603    & 182       & 6     & 40    & 2.262   & 0.180    & 7.21                   & 140   \cr
SU(3) & 0.603    & 167       & 6     & 30    & 5.642   & 0.197    & 5.91                   & 62    \cr
\hline
SU(2) & 0.733    & 222       & 6     & 40    & 2.332   & 0.148    & 5.93                   & 326   \cr
SU(3) & 0.733    & 203       & 6     & 30    & 5.738   & 0.162    & 4.86                   & 64    \cr
\hline
SU(2) & 0.903    & 273       & 4     & 46    & 2.2615  & 0.181    & 8.30                   & 246   \cr
SU(3) & 0.903    & 250       & 4     & 34    & 5.642   & 0.197    & 6.70                   & 59    \cr
\hline
SU(2) & 0.968    & 293       & 4     & 46    & 2.2872  & 0.168    & 7.75                   & 256   \cr
SU(3) & 0.968    & 268       & 4     & 34    & 5.675   & 0.184    & 6.16                   & 63    \cr
\hline
SU(2) & 0.986    & 298       & 4     & 46    & 2.2938  & 0.165    & 7.61                   & 235   \cr
SU(3) & 0.986    & 273       & 4     & 34    & 5.685   & 0.180    & 6.12                   & 66    \cr
\hline
SU(2) & 1.00     & 303       & 4     & 46    & 2.299   & 0.163    & 7.50                   & 292   \cr
SU(3) & 1.00     & 277       & 4     & 34    & 5.69236 & 0.178    & 6.05                   & 65    \cr
\hline
SU(2) & 1.02     & 308       & 4     & 46    & 2.306   & 0.160    & 7.35                   & 280   \cr
SU(3) & 1.02     & 282       & 4     & 34    & 5.7     & 0.175    & 5.95                   & 67    \cr
\hline
SU(2) & 1.04     & 314       & 4     & 46    & 2.313   & 0.157    & 7.21                   & 319   \cr
SU(3) & 1.04     & 288       & 4     & 34    & 5.71    & 0.171    & 5.81                   & 113   \cr
\hline
SU(2) & 1.10     & 333       & 4     & 46    & 2.333   & 0.148    & 6.80                   & 276   \cr
SU(3) & 1.10     & 305       & 4     & 34    & 5.738   & 0.162    & 5.51                   & 71    \cr
\hline
SU(2) & 1.81     & 548       & 2     & 52    & 2.263   & 0.180    & 9.35                   & 232   \cr
SU(3) & 1.81     & 500       & 2     & 38    & 5.642   & 0.197    & 7.49                   & 59    \cr
\hline
SU(2) & 2.20     & 665       & 2     & 52    & 2.332   & 0.148    & 7.71                   & 310   \cr 
SU(3) & 2.20     & 609       & 2     & 38    & 5.738   & 0.162    & 6.16                   & 70    \cr 
\hline 
\end{tabular}
\end{center} 
\end{table*}

\begin{appendix}

\section{Simulation parameters and volume effects}\label{app:simul}

In table \ref{conf} we present the simulation parameters for the lattice calculation of the 
temperature dependent gluon propagator. The scale for SU(2) and SU(3) has been 
fixed according to \cite{Fingberg:1992ju} and \cite{Lucini:2003zr}, respectively, 
setting the string tension to $\sigma=(440 $MeV$)^2$, to which the 
considerations of \cite{Maas:2007af} apply. $T_c$ is then 303 MeV for SU(2) 
and 277 MeV for SU(3), at $\beta=2.299$ and $\beta=5.69236$, respectively, for 
the employed value of $N_t=4$ at $T_c$. $T$ is the corresponding temperature, 
$N_t$ the temporal extent of the lattice, $N_s$ is the spatial extent, 
$\beta$ the bare coupling, $a$ the lattice spacing and $V_s$ the corresponding 
spatial physical volume. The number of configurations is denoted by conf., and 
the number of thermalization sweeps is given by $200+10N_s$, and of decorrelation 
sweeps by $20+N_s$. Use has been made of the fact that the larger the number of 
generators, the less statistics is needed for the gluon propagator, due to the 
generator-averaging. The statistics was aimed at below the ten-percent 1$\sigma$ 
statistical error level for the electric screening mass.

\begin{figure*}
\includegraphics[width=0.4\linewidth]{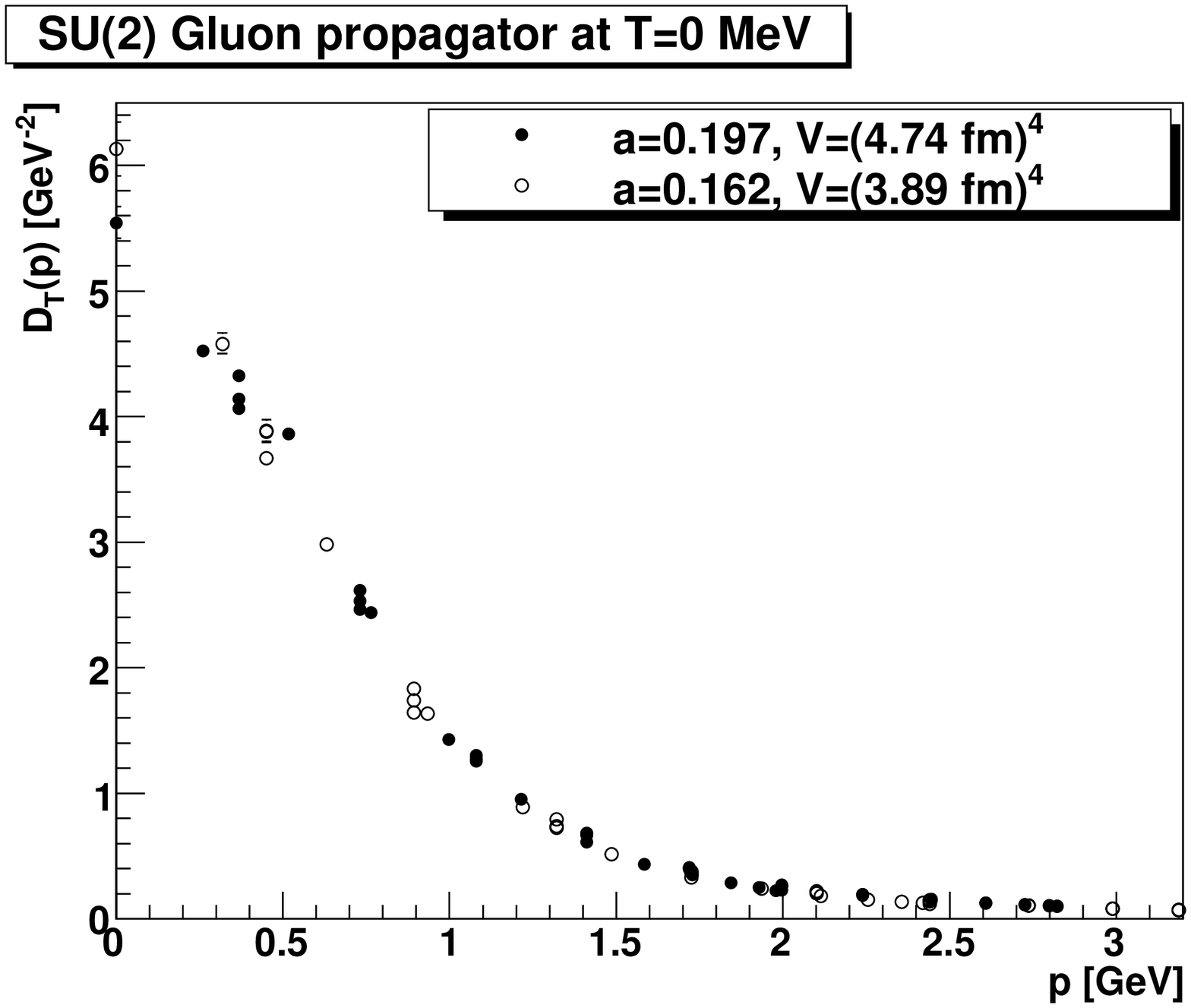}\hfill\includegraphics[width=0.4\linewidth]{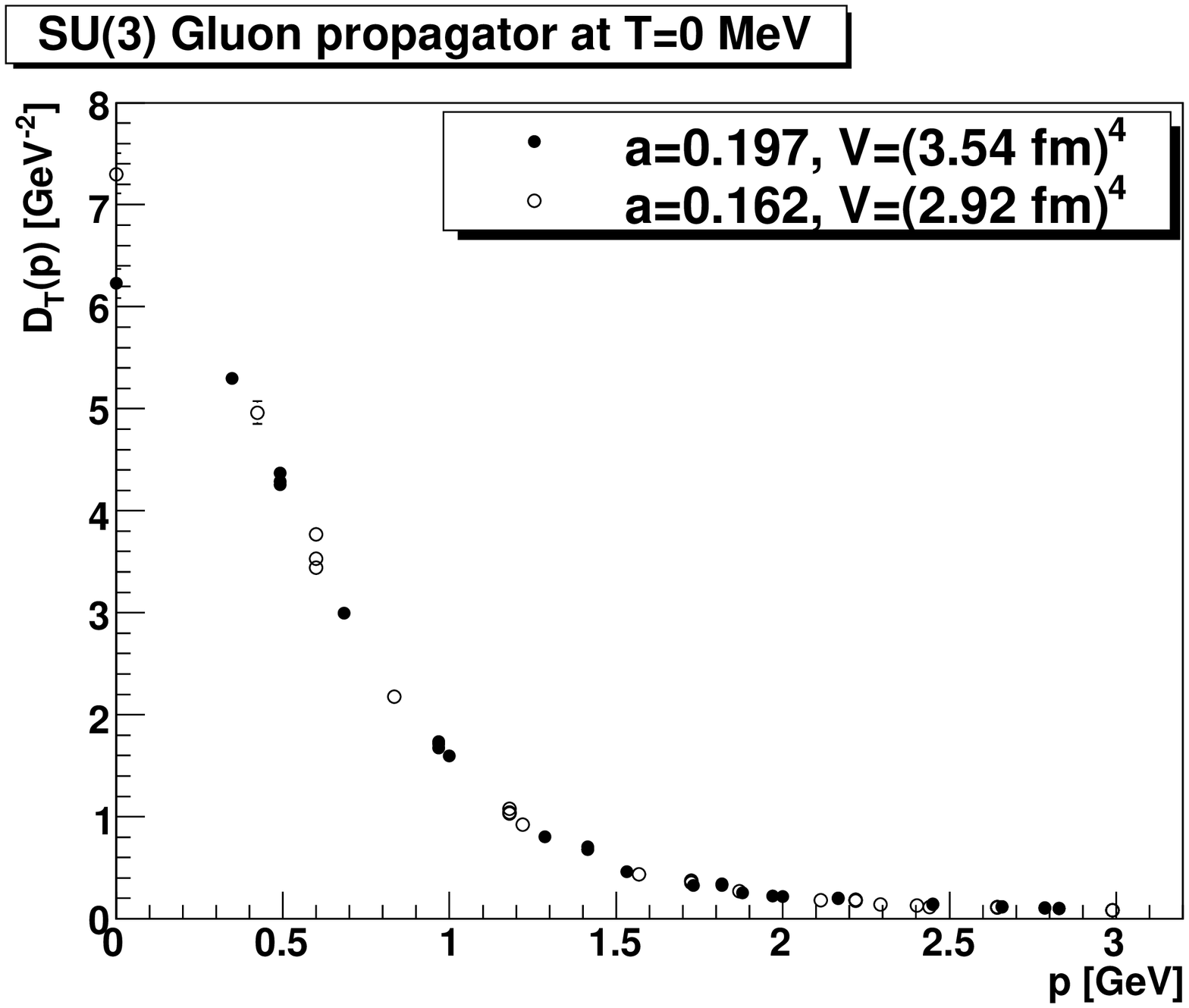}
\caption{\label{lat-gpt-t0}The gluon propagator at zero temperature. SU(2) is shown in the left panel and SU(3) in the right panel.}
\end{figure*}
\begin{figure*}
\includegraphics[width=0.4\linewidth]{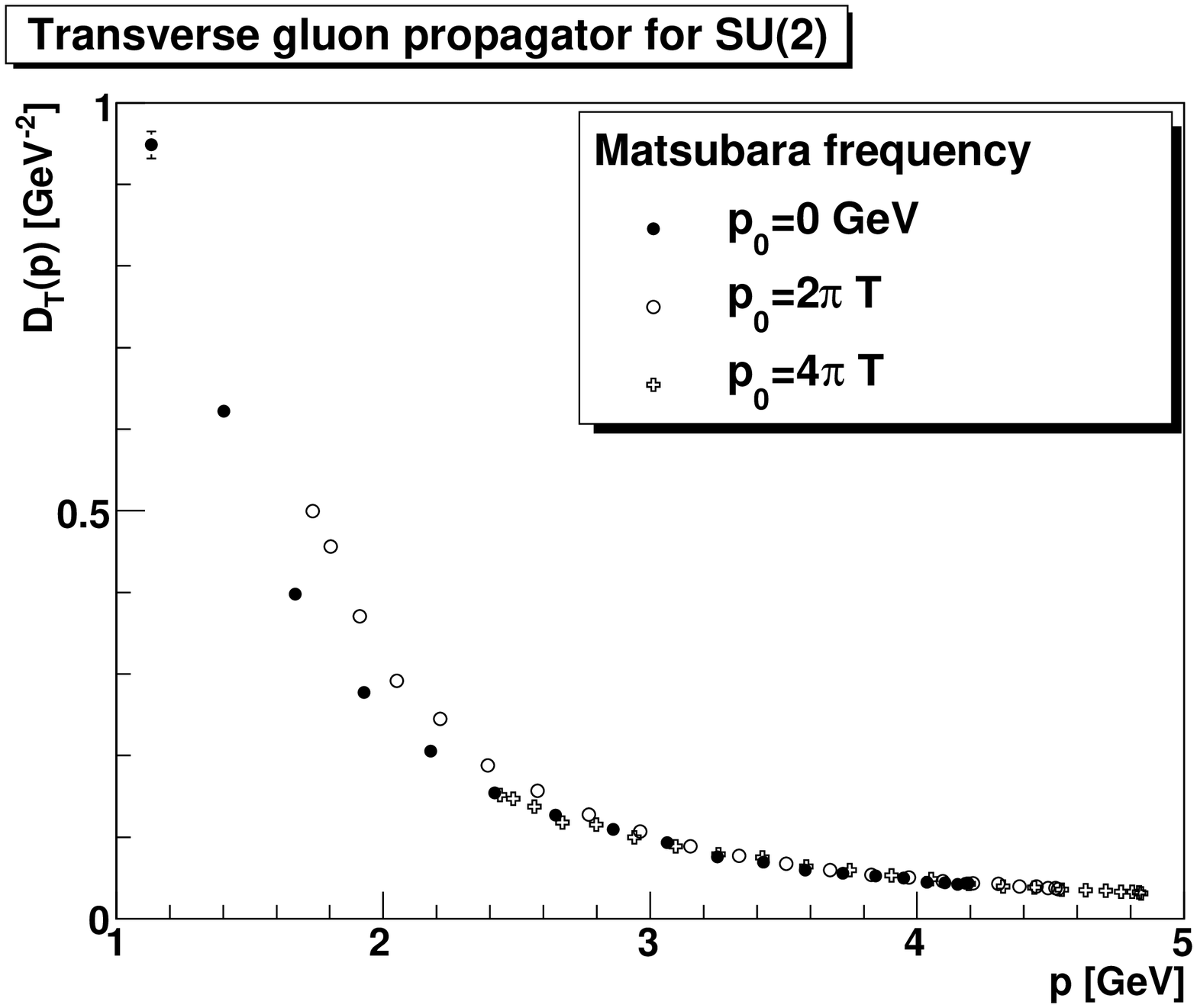}\hfill\includegraphics[width=0.4\linewidth]{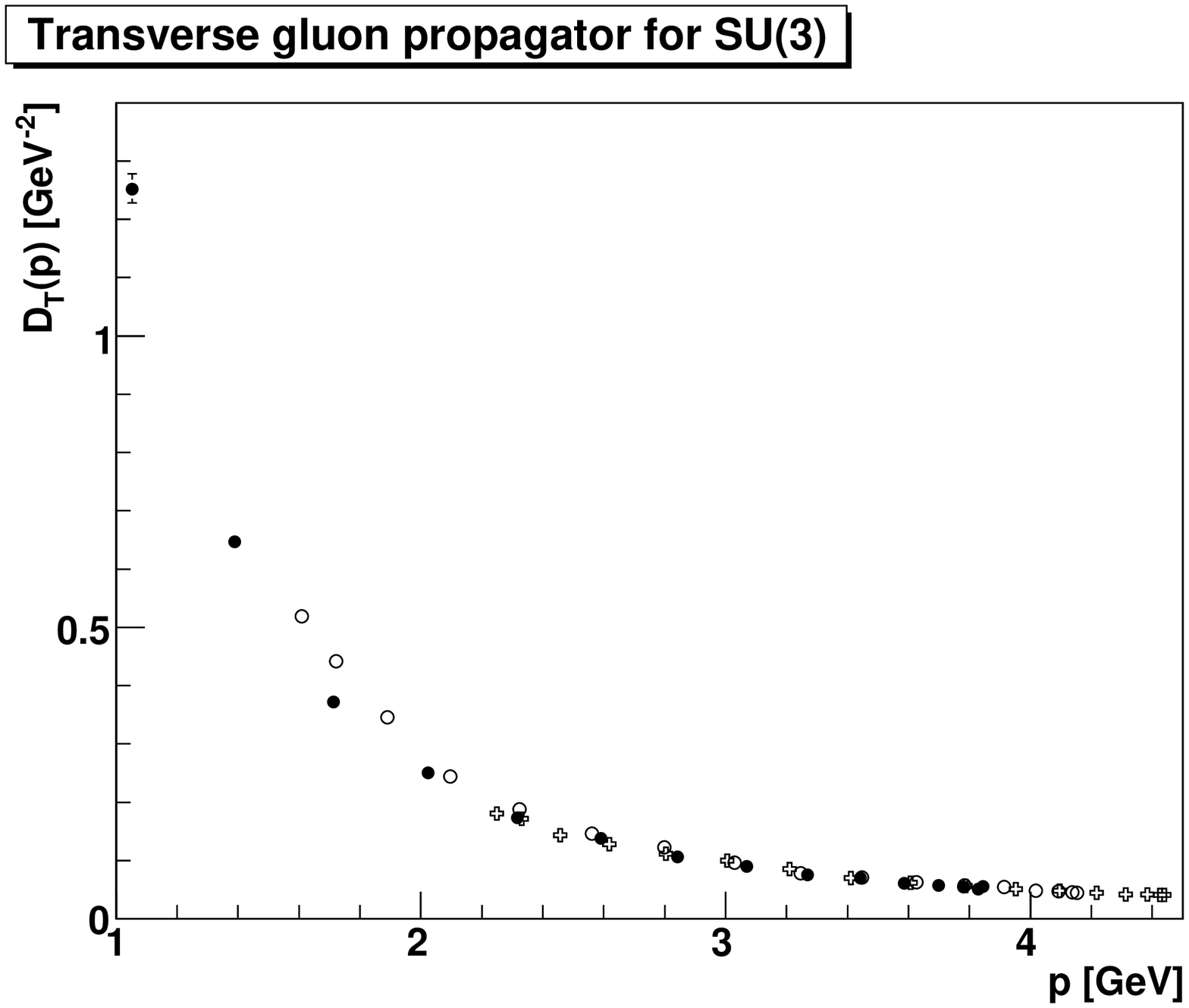}\\
\includegraphics[width=0.4\linewidth]{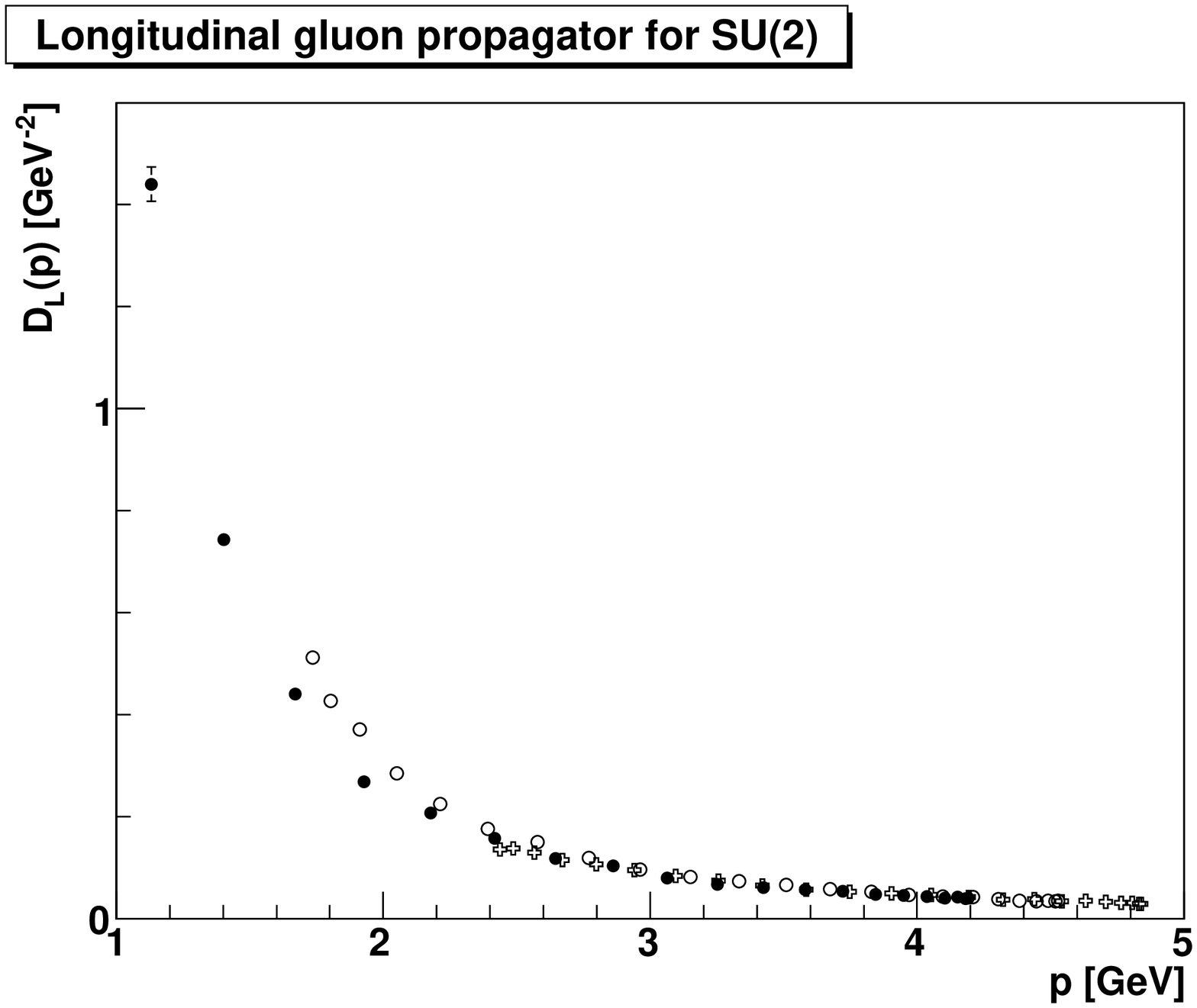}\hfill\includegraphics[width=0.4\linewidth]{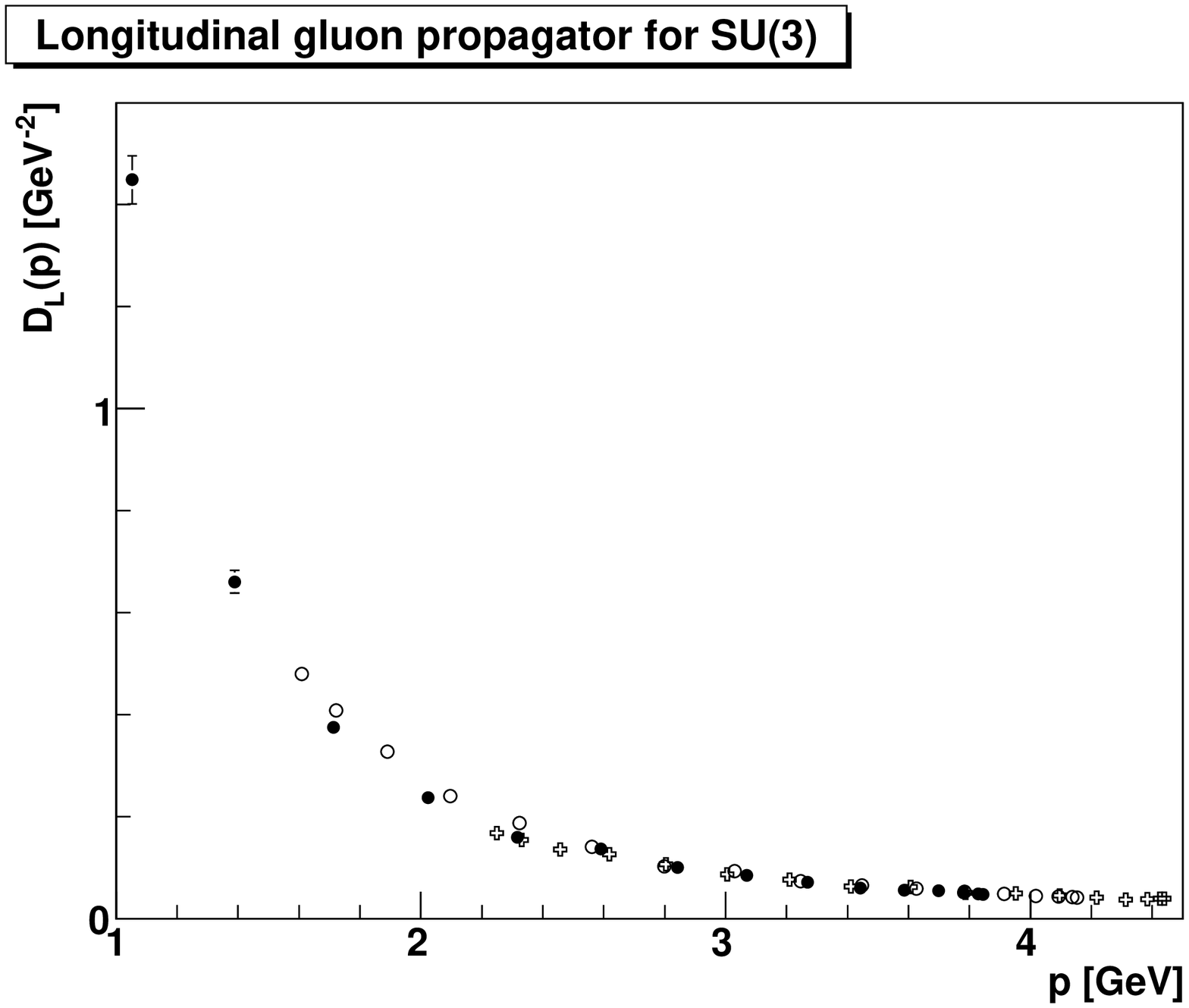}
\caption{\label{lat-hard}Comparison of the soft mode to the first two higher Matsubara frequencies at $T=T_c$. In the upper panels the transverse gluon propagator is shown, and in the lower panel the longitudinal ones. In the left panels the results for SU(2) are compared to the results for SU(3) in the right panels. All results are shown as a function of the four momentum $p^2=\op^2+(\vec p)^2$.}
\end{figure*}
\begin{figure*}[htb]
\includegraphics[width=0.4\linewidth]{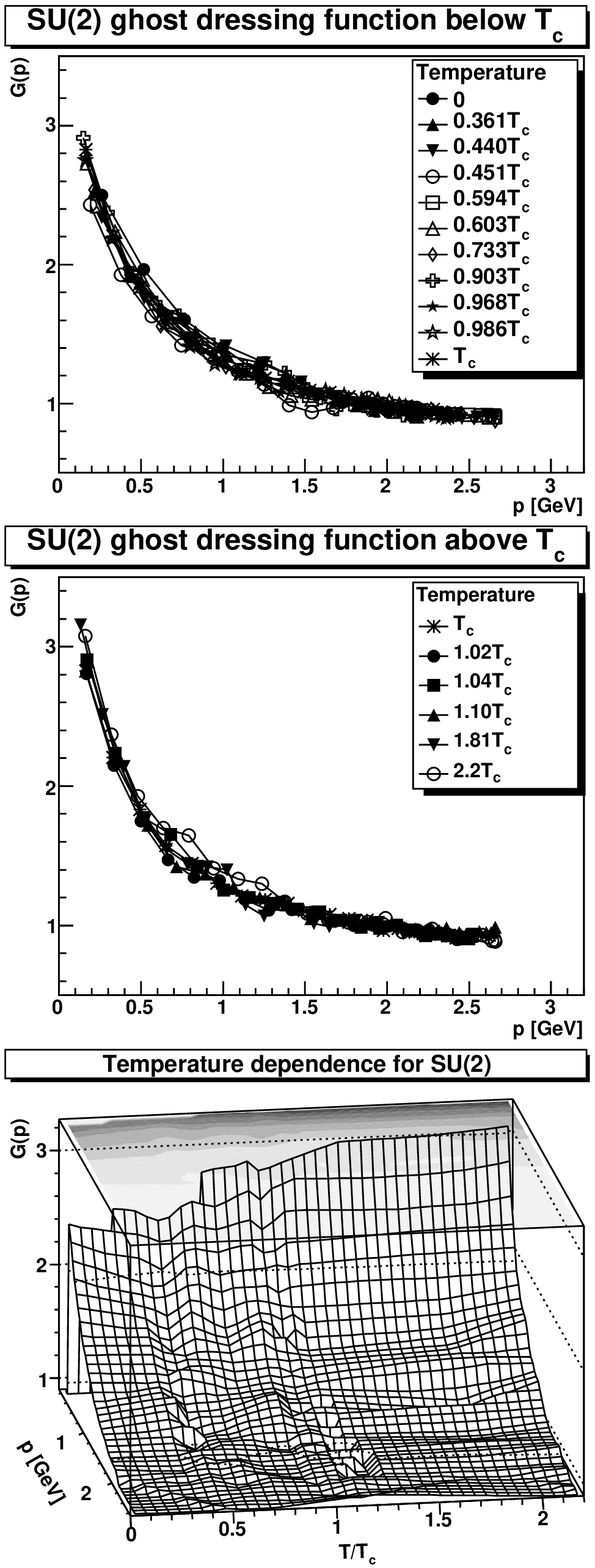}\hfill\includegraphics[width=0.4\linewidth]{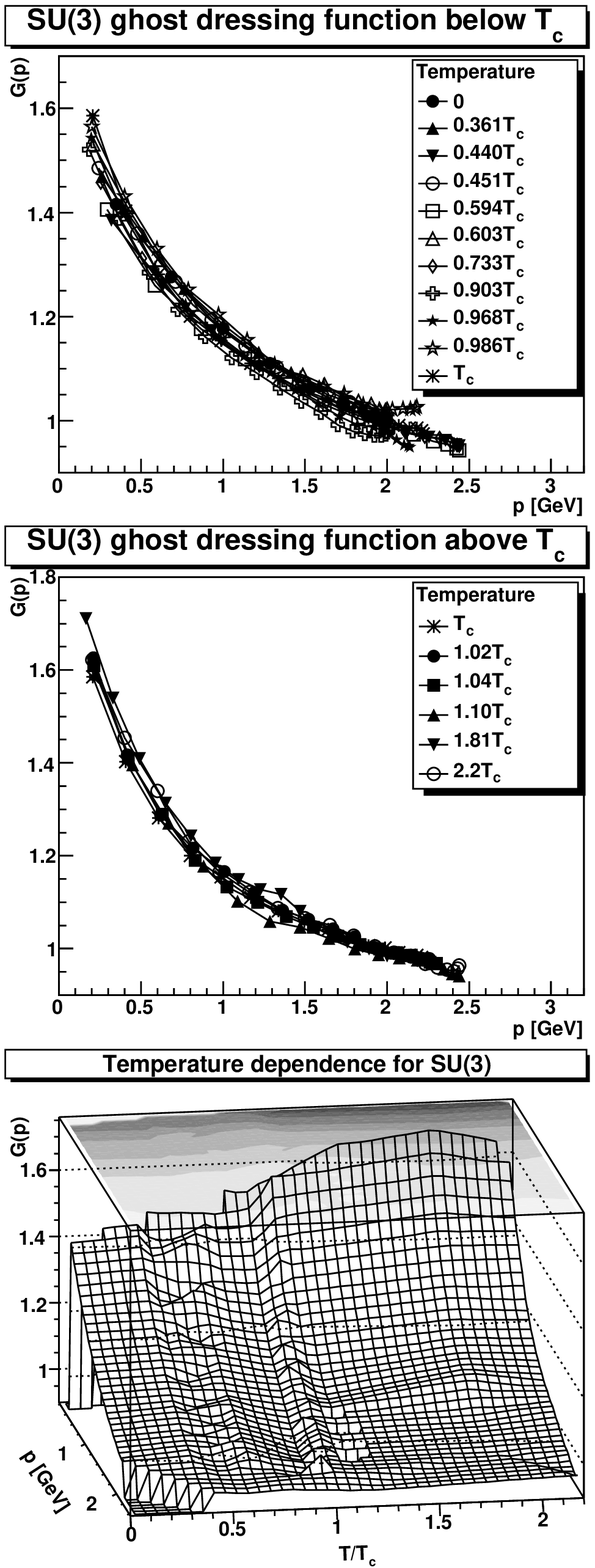}
\caption{\label{lat-ghp}The ghost dressing function $G(p)=p^2 D_G(p)$ of the ghost propagator $D_G$ at temperatures below (top panel) and above (middle panel) the phase transition. The bottom panel shows both results together. On the left results for SU(2) and on the right results for SU(3) are shown. Lines are drawn to guide the eye.}
\end{figure*}

An immanent drawback 
for the comparison of SU(2) to SU(3) is the lower phase transition temperature of SU(3) 
in physical units. As a consequence, the already more expensive calculation for SU(3) 
would become even more expensive if the same physical volumes in spatial direction at 
a fixed time-extension would be required for SU(3). In this first investigation for 
SU(3), this has not been done, at the expense of larger finite-volume artifacts at low 
momenta for SU(3). These are exemplified in figure \ref{lat-gpt-t0}, where the results 
for the zero-temperature propagator for the smallest and largest discretization for SU(3) and SU(2)
are compared, as an estimate for the systematic uncertainties involved. At finite 
temperature, the spatial volumes are larger, but in particular for the chromoelectric 
gluon this is counteracted by discretization artifacts \cite{Cucchieri:2007ta}.

\section{Soft vs. hard modes}\label{app:hard}

At the temperatures studied here, functional studies suggest that the higher 
Matsubara frequencies can be approximated rather well by 
$D_T(\op^2,\vec p^2)=D_T(0,\op^2+\vec p^2)$, and likewise for the other propagators 
\cite{Cucchieri:2007ta,Maas:2005hs}. That this is indeed a rather good approximation for the 
volumes employed here is shown in figure \ref{lat-hard}.
\begin{table*}[t]
\caption{Temperature dependent fit parameters for the SU(2) and SU(3) gluon propagator.\label{tab1}}
\begin{tabular}{c||c|c|c|c|c|c|c|c|c|c|c|c|c|c|c|c}
$SU(2)$      & & & & &  &  & && 	&&& 	&	& 	&\\\hline\hline
$T/T_c$      & 0    & 0.361& 0.44  & 0.451 & 0.549 & 0.603 &0.733 &0.903&0.968 	&0.986 	&1 &1.02& 1.04 	&1.1 	&1.81 	&2.2\\\hline\hline
$a_{L}(T)$   & 1.22 & 0.62& 0.48 & 0.63 & 0.30 & 0.29 &0.21 &0.15&0.15	&0.15	&0.18	&0.34	&0.56	&0.93	&2.69	&5.00\\   
$b_{L}(T)$   & 1.94& 1.48& 1.41 & 1.52 & 1.31 & 1.29 &1.29 &1.28&1.30	&1.29	&1.28	&1.31	&1.41	&1.56	&1.15	&1.16\\ \hline  
$a_{T}(L)$   & 1.22 & 1.31& 1.31 & 1.39 & 1.31 & 1.32   &1.32 &1.33&1.32&1.29	&1.24	&1.26	&1.30	&1.37	&1.33	&1.35	\\  
$b_{T}(L)$   & 1.94& 1.98 & 2.02 & 2.03 & 1.98 &1.92  &1.89 &1.76&1.75	&1.72	&1.71	&1.76   &1.79	&1.83	&1.48	&1.32\\\hline\hline 
$SU(3)$      & & & & &  &  & && 	&&& 	&	& 	&\\\hline\hline
$T/T_c$      &0	  & 0.361& 0.44  & 0.451 & 0.549 & 0.603 &0.733 &0.903	&0.968 	&0.986 	&1 	&1.02	& 1.04 	&1.1 	&1.81 	&2.2\\\hline\hline
$a_{L}(T)$   &0.60&0.42&0.23	& 0.33 & 0.19	&0.17	&0.11	&0.098	&0.082	&0.079	&0.16	&0.27	&0.32	&0.50	&2.71	&4.72\\
$b_{L}(T)$   &1.36&1.23&1.14	& 1.20 & 1.13	&1.08	&1.10	&1.13	&1.14	&1.14	&1.05	&1.05	&1.03	&1.07	&1.14	&1.47\\\hline
$a_{T}(T)$   &0.60&0.71&0.78	& 0.83 & 0.86	&1.04	&1.05	&1.67	&1.57	&1.06	&0.54	&0.55	&0.57	&0.63	&1.47	&1.42\\
$b_{T}(T)$   &1.36&1.37&1.46	& 1.47 & 1.52	&1.60	&1.60	&1.91	&1.81	&1.45	&1.13	&1.14	&1.17	&1.19	&1.49	&1.30\\\hline\hline
\end{tabular}
\end{table*}

\begin{figure*}[t]
\includegraphics[width=0.9\columnwidth]{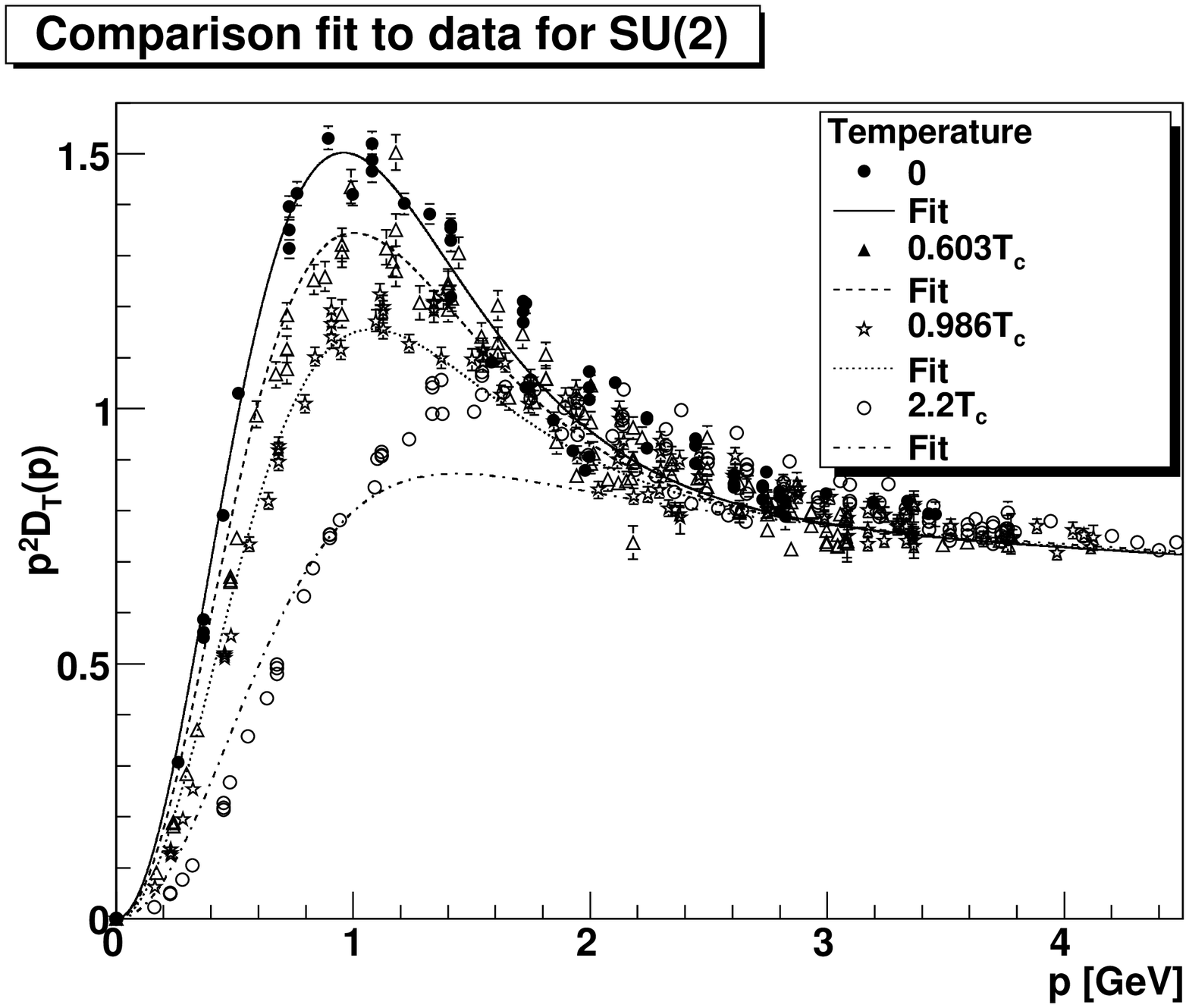}\hfill
\includegraphics[width=0.9\columnwidth]{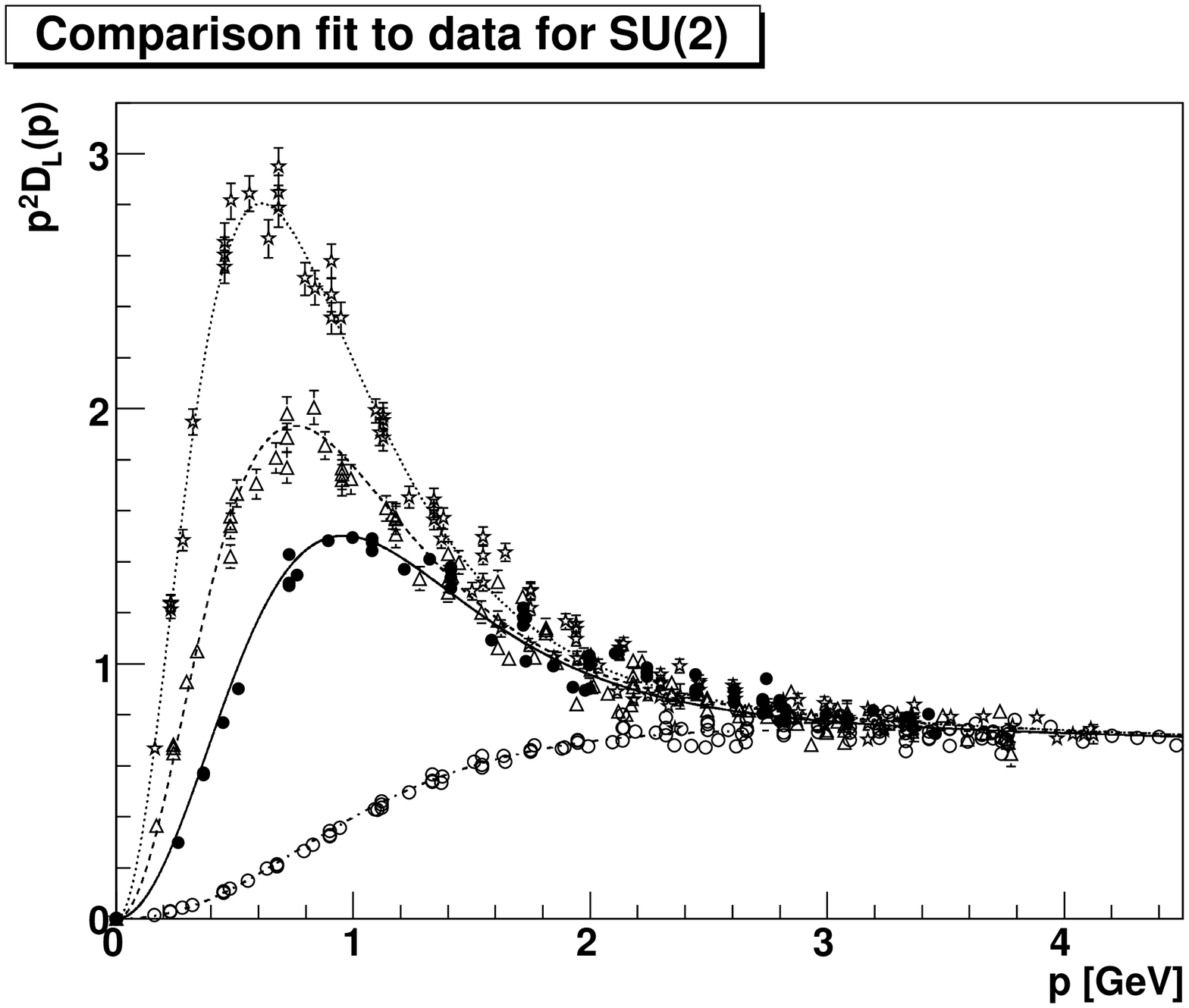}
\caption{
Magnetic (left) and electric (right) temperature dependent gluon
dressing function from our lattice calculations compared with our
fits.}
\label{res:fitscompared}
\end{figure*}

\section{Ghost dressing function}\label{app:ghost}

The soft mode of the ghost is also associated with the chromomagnetic sector of the 
theory \cite{Cucchieri:2007ta,Maas:2005hs,Maas:2004se}. Consequently, in accordance 
with the expectation \cite{Cucchieri:2007ta}, it shows essentially no dependence on 
the temperature, see figure \ref{lat-ghp}. If the ghost is indeed dominated by the 
gauge-fixing procedure this shows that the procedure is not affected significantly 
by a thermodynamic environment. The only observable effect, an increase in infrared 
strength with increasing temperature, is likely only an effect due to the larger 
spatial volumes used at higher temperatures.

\section{Fit functions for the temperature-dependent gluon propagator}\label{app:fit}

As already detailed in the main body of this work around \Eq{gluefit} we use a
fit function for the temperature dependent longitudinal and transverse parts
of the gluon propagator to represent the corresponding lattice data. 

The resulting fit parameters $a_{T,L}(T)$ and $b_{T,L}(T)$ for the gauge
groups SU(2) and SU(3) are given in tables \ref{tab1}. The rather good quality of the fits, given the statistical accuracy and systematic errors of the lattice calculations, can be seen in figures \ref{res:fits} and \ref{res:fitscompared}.

\end{appendix}

\end{document}